\documentclass[11pt,a4paper]{article}
\usepackage{jcappub}
\usepackage{graphicx}
\usepackage[utf8]{inputenc}
\usepackage[T1]{fontenc}
\usepackage[usenames,dvipsnames]{xcolor}
\usepackage{hyperref}
\hypersetup{colorlinks=true,
citecolor=ForestGreen, linkcolor=RoyalBlue, urlcolor=Turquoise}
\usepackage{tabularx,rotating}
\usepackage{subcaption}
\usepackage{float}
\usepackage{orcidlink}

\DeclareUnicodeCharacter{2212}{\ensuremath{-}}

\allowdisplaybreaks

\title{Alleviating the $H_0$ and $\sigma_8$ tensions in the interacting cubic covariant Galileon model}

\author[]{Sihem Zabat, }	
\author[]{Youcef Kehal \orcidlink{0009-0004-9312-720X},}
\author[]{and Khireddine Nouicer \orcidlink{0000-0001-6110-9833}}

\affiliation[]{Laboratory of Theoretical Physics and Department of Physics,\\ Faculty of Exact and Computer Sciences, University of Jijel,\\ B.P. 98, Ouled Aissa, Jijel 18000, Algeria}

\emailAdd{sihem.zabat@univ-jijel.dz}
\emailAdd{youcef.kehal@univ-jijel.dz}
\emailAdd{khnouicer@univ-jijel.dz}

\date{\today}
	
	\abstract{
		The interaction between dark matter and dark energy has become a focal point in contemporary cosmological research, particularly in addressing current cosmological tensions. This study explores the cubic Galileon model's interaction with dark matter, where the interaction potential in the dark sector is proportional to the dark energy density of the Galileon field. By employing dimensionless variables, we transform the field equations into an autonomous dynamical system. We calculate the critical points of the corresponding autonomous systems and demonstrate the existence of a stable de Sitter epoch.
Our investigation proceeds in two phases. First, we conduct a detailed analysis of the exact interacting cubic Galileon (ICG) model, derived from the precise solution of the equations of motion. Second, we explore an approximate tracker solution, labeled TICG, assuming a small coupling parameter between dark matter and dark energy. We evaluate the evolution of these models using data from two experiments, aiming to resolve the tensions surrounding $H_0$ and $S_8$.
The analysis of the TICG model indicates a preference for a phantom regime and provides a negative coupling parameter in the dark sector at a $68\%$ confidence level. This model also shows that the current tensions regarding $H_0$ and $S_8$ are alleviated. Conversely, the ICG model, despite its preference for the phantom regime, is plagued by an excess in today's matter density and a higher expansion rate, easing only the $H_0$ tension.}
\keywords{DE-DM interaction, cubic Galileon model, $H_0$ and $S_8$ tensions}
	\arxivnumber{2305.12155}

\begin{document}
\maketitle

	%%%%%%%%%%%%%%%%%%%%%%
\section{Introduction}

	It goes without saying that the concordance $\Lambda$CDM model stands as the prevailing framework used in cosmology to describe the large-scale structure and evolution of the universe. It postulates the existence of a non-baryon cold dark matter (DM) alongside ordinary  matter, playing the essential role in explaining the rotation curves of galaxies and the processes of structures formation \cite{bertone2005particle,Bertone:2010zza}. Additionally, the $ \Lambda $CDM model aligns consistently with a wide range of observational data, including the current accelerated expansion of the universe \cite{riess1998observational,perlmutter1999measurements,spergel2003first,hinshaw2013nine,aghanim2020planck}. This latter, can be explained in the framework of general relativity (GR) by introducing an exotic fluid with negative pressure known as dark energy (DE) \cite{copeland2006dynamics}, which is often linked to the cosmological constant  \cite{carroll2001cosmological}. Despite its success, the model still face unresolved issues, casting doubt on its validity \cite{bull2016beyond,perivolaropoulos2022challenges} such as the cosmological constant problem and the coincidence problem. The first one is related to the huge discrepancy between the theoretical value of vacuum energy density $ \Lambda $ predicted by quantum field theories and the experimental value of DE in the universe \cite{Wein} while the second one revolves around the peculiar observation that the current densities of vacuum energy  and matter in the universe are roughly comparable, despite their distinct evolutionary behaviors \cite{velten2014aspects}.
	
	In this regard, alternative directions have been followed to address these problems and demystify the origin of DE that can reproduce the late-time acceleration of the universe. A common way consists of modeling the right hand side of Einstein's equations with specific forms of matter \cite{tsujikawa2011dark}. For instance, scalar fields can be naturally introduced as candidates for DE such as quintessence model \cite{fujii,ford,WETTERICH1988668,caldwell1998cosmological,zlatev1999quintessence} in which the quintessence field is minimally coupled to gravity and its equation of state (EoS) is no longer a constant but it rather slightly evolves in time \cite{tsujikawa2013quintessence,copeland2006dynamics}. Other dynamical DE models have been proposed including k-essence \cite{armendariz2000dynamical}, tachyons \cite{padmanabhan2002accelerated,padmanabhan2002can} and Chaplygin gas \cite{kamenshchik2001alternative,bilic2002unification,bento2002generalized}. On the other hand, a different path postulates that GR is only accurate at local systems and fails to describe the universe at larger scales and hence it should be modified \cite{clifton2012modified,nojiri2017modified,saridakis2021modified}. Scalar-tensor theories \cite{quiros2019selected} represent a compelling approach within the realm of modified gravity to elucidate the mysteries surrounding DE and cosmic inflation. These theories encompass a broader framework than GR incorporating scalar fields as additional degree of freedom alongside the metric. One prominent example is the Galileon models \cite{nicolis2009galileon} which are a subset of Horndeski theories, and leads to second order field equations even though their Lagrangians contain higher order derivatives of the field \cite{kobayashi2019horndeski,deffayet2013formal}. Galileon is a scalar field $ \pi $ that satisfies the Galilean shift symmetry $\partial_{\mu}\pi\rightarrow\partial_{\mu}\pi+b_{\mu}$ in flat space-time \cite{kobayashi2019horndeski}, while this property is broken in the covariant version \cite{deffayet2009covariant,deffayet2009generalized} when the Lagrangian is extended to curved space-time. Emerging from the decoupling limit of the DGP braneworld \cite{dvali20004d,luty2003strong,nicolis2004classical} to avoid ghost modes, the covariant Galileon field and its extended models have sparked extensive research into their phenomenological aspects. This exploration aims to probe their roles in producing a self-accelerating phase and primordial inflation \cite{kobayashi2010inflation,burrage2011galileon,deffayet2010imperfect,chow2009galileon,silva2009self,kobayashi2010cosmic}, or even proposing a self tuning mechanism to alleviate the cosmological constant problem \cite{charmousis2012general,copeland2012cosmology}. Galileon theories are usually relevant to the Vainshtein mechanism, where the scalar mode is weakly coupled to the source. Since that the presence of non-linear derivative interactions in the scalar sector allows to screen the extra force mediated by the scalar field on small scales. 	
	
	Apart from the theoretical challenges mentioned above, another set of problems emerges from cosmological and astrophysical data. In particular, the present value of the Hubble parameter $ H_0 $ is estimated by the 	Cosmic Microwave Background (CMB) constraints and the Planck collaboration as $ H_0=67.36 \pm 0.54$ km/s/Mpc \cite{aghanim2020planck}, while local distance ladder measurements from Type Ia supernovae from the 2019 local measurements by SH0ES collaboration	 (R19) indicate $ H_0=(73.04 \pm 1.04) $ km/s/Mpc \cite{Riess_2022,2023JCAP} reporting a significant tension of $\sim 5 \sigma $. An other tension within the concordance model is the inconsistency between CMB and LSS observations, quantified in terms of the $S_8 = \sigma_8 \sqrt{\Omega_m^{(0)} /0.3}$ parameter, where $\Omega_m^{(0)}$ is the present day  matter density. Indeed, the Dark Energy Survey Year-3 (DES-Y3) \cite{2022PhRvD.105b3520A} have revealed a $2.5\sigma$ tensions with   Planck data assuming $\Lambda $CDM model. These tensions, appear not attributed to unknown systematic effects but may signal a potential breakdown of the $ \Lambda $CDM model. In the same spirit, non-gravitationally interacting models between DE and DM were originally introduced to justify the cosmological constant and coincidence problems \cite{wang2016dark}, and they also seem to be effective in alleviating the $ H_{0} $ \cite{di2017can,buen2018interacting,yang2018tale,yang2018interacting,pan2019observational,yang2019dark,pan2019interacting,martinelli2019constraints,pan2020non,yang20212021,yang2023dynamics,kumar2019dark} and $ \sigma_8 $  \cite{kumar2019dark,pourtsidou2016reconciling,an2018relieving,van2018searching} tensions.
	Inspired by recent approaches, the primary objective of this paper is to explore potential solutions or mitigation for the cosmological tensions involving $H_0$ and $S_8$ by incorporating interacting dark energy (DE) with dark matter (DM) within the framework of the cubic covariant Galileon model. We introduce an interaction term expressed as $Q=\alpha H \rho_{x}$, where $H$ represents the Hubble parameter,  $\rho_x$ denotes the energy density of the Galileon field, and $\alpha$ is a dimensionless coupling parameter.

This paper is organized as follows. In Sec. \ref{sec:Cosmological-Galileon}, we introduce the Interacting Cubic Covariant Galileon (ICG) model and derive the corresponding field equations, assuming a spatially flat FLRW space-time. We also analyze the existence and stability of fixed points and extract constraints on the DM-DE coupling constant. Sec. \ref{sec:Growth-rate-of} is dedicated to studying the evolution of cosmological perturbations in the presence of perfect fluid matter, investigating the behavior of the growth rate of matter perturbations and the gravitational potential in the quasi-static approximation on sub-horizon scales. In Sec. \ref{sec:Methodology and data}, we describe the cosmological data and the methodology used to obtain constraints on model parameters. Sec. \ref{sec:results and discussion} presents the cosmological constraints on model parameters for both the exact and tracker solutions using a Monte Carlo Markov Chain (MCMC) approach. We compare these results with the $\Lambda$CDM model and use the corrected frequentist Akaike Information Criterion ($\textrm{AIC}_\textrm{c}$) to assess whether the ICG model is favored over the $\Lambda$CDM model. Finally, our conclusions are presented in Sec. \ref{sec:Conclusion}.
	\section{ICG on the background \label{sec:Cosmological-Galileon}}
	
The action of minimally coupled cubic covariant Galileon field is described
	by the action \cite{deffayet2009generalized}
\begin{equation}
		S=\int d^{4}x\sqrt{-g}\left(\frac{M_{\textrm{Pl}}^{2}R}{2}+\frac{1}{2}c_{2}\pi_{;\mu}\pi^{;\mu}+\frac{1}{2}\frac{c_3}{M^3}\left(\nabla\pi\right)^{2}\square\pi+\mathcal{L}_{M}\left[g_{\mu\nu},\left\{ \phi_{a}\right\} \right]\right)\label{Action}
\end{equation}
where $M_{\textrm{Pl}}^{2}$ is the reduced Planck mass, $M_{\textrm{Pl}}^{2}=1/8\pi G$,  $c_{i}$ are constants
	parameters, $M$ is a constant with dimensions of mass, and $\mathcal{L}_{M}$ is the matter Lagrangian. 
	 
	Varying the action ($\ref{Action}$) with respect to $g_{\mu\nu}$ we obtain  the Einstein's field  equations: 
	\begin{equation}
		M_{\textrm{Pl}}^{2}G_{\alpha\beta}=\sum_{i=2}^{3}c_{i}T_{\alpha\beta}^{\left(i\right)}+\sum_{A}T_{\alpha\beta}^{\left(a\right)},\label{ten2}
	\end{equation}
	where $G_{\alpha\beta}$ denotes the Einstein symmetric tensor,  and  \begin{equation}
		T_{\alpha\beta}^{\left(2\right)}=-\pi_{;\alpha}\pi_{;\beta}+\frac{1}{2}g_{\alpha\beta}\left(\nabla\pi\right)^{2}\text{ },\label{eq:t-2}
\end{equation}
\begin{equation}
		T_{\alpha\beta}^{\left(3\right)}=-\frac{1}{M^{3}}\left(\pi_{;\alpha}\pi_{:\beta}\square\pi+g_{\alpha\beta}\pi_{;\mu}\pi^{;\mu\lambda}\pi_{;\lambda}-\pi^{;\mu}\left[\pi_{;\alpha}\pi_{;\beta\mu}+\pi_{;\beta}\pi_{;\alpha\mu}\right]\right),\label{eq:t-3}
	\end{equation}
	are the contributions to the Galileon energy-momentum tensor, 
and $T_{\alpha\beta}^{\left(A\right)}$ is
	the energy-momentum tensor of radiation, baryon matter and dark
	matter $\left(A=\text{r,\:b\:,c}\right)$. 
	For matter components, we assume the usual energy-momentum tensor
	describing a perfect fluid 
	\begin{equation}
		T_{\alpha\beta}^{\left(A\right)}=\left(\rho_{A}+p_{A}\right)u_{\alpha}^{A}u_{\beta}^{A}+p_{A}g_{\alpha\beta},\label{eq:Perfect_Fluid}
	\end{equation}
	where $\rho_{A}$ is the energy density, $p_{A}$ the pressure, and
	$u_{\alpha}^A$ the velocity of the $A$-fluid. 
	In order to preserve the local energy-momentum conservation law, the Bianchi identities imply that
	\begin{equation}
		\nabla_{\alpha}T^{\alpha\beta}=0.\label{eq:Cons_tot}
	\end{equation}
	In the absence of coupling between matter and energy species we have $\nabla_{\alpha}{T^{\alpha}}_{\beta}^{(A)}=0$, which  is the case for standard model of particles \cite{carroll1998quintessence}.
	In order to study the covariant Galileon model in its generality and
	allow to the existence of energy transfer between DM
	and DE  as supported by observations of galaxy clusters
	\cite{bertolami2007dark}, we assume that the dark sectors
	do not evolve separately but interact with each other. The usual
	way to describe this interaction is to introduce an energy-momentum
	exchange current into the conservation equations as follows 
	\begin{equation}
		\nabla_{\alpha}T_{\text{x}}^{\alpha\beta}=-\nabla_{\alpha}T_{\text{c}}^{\alpha\beta}=Q^{\beta},\label{eq:interaction}
	\end{equation}
	where $Q^{\beta}$ is given covariantly by (see \cite{valiviita2008large} and references therein)
	\begin{equation}
		Q_{\text{{x}/c}}^{\beta}=Qu^{\beta},
	\end{equation}
	$u^{\beta}$ is the DE/DM 4-velocity and $Q$ is 
	the interaction function between DE and DM, and generally it is
	a function of DE and DM densities, the Hubble parameter and its derivatives.
	Assuming that there is only energy transfer between DE and DM we have
	$Q_{x}=-Q_{\text{c}}=Q.$ In this study we are interested by the case $Q \sim \rho_{x}$, where positive $Q$ indicates
	that DM decays to DE, whereas DE decays to DM for negative $Q$.
\subsection{Dynamical system analysis}\label{dyn.sys}
	To study the background cosmological dynamics of the ICG, we assume the geometry of spatially flat expanding
	universe described by the FLRW metric 
	\begin{equation}
		ds^{2}=-n^{2}\left(t\right)dt^{2}+a^{2}(t)\delta_{ij}dx^{\text{i}}dx^{j}
	\end{equation}
	where $a\left(t\right)$ is the scale factor. Using $n^{2}\left(t\right)=1,$
	the Friedmann equations on the FLRW background are obtained from the
	$(0,0)$ and $(i,j)$ components of Einstein equations ($\ref{ten2})$
	\begin{equation}
		3M_{\text{Pl}}^{2}H^{2}=-\frac{1}{2}c_{2}\dot{\pi}^{2}+\frac{3}{M^{3}}c_{3}H\dot{\pi}^{3}+\rho_{\text{c}}+\rho_{\text{b}}+\rho_{\text{r}}\text{ },\label{f1}
	\end{equation}
	\begin{equation}
		M_{\text{Pl}}^{2}(3H^{2}+2\dot{H})=\frac{1}{2}c_{2}\dot{\pi}^{2}+\frac{1}{M^{3}}c_{3}\dot{\pi}^{2}\ddot{\pi}-\frac{1}{3}\rho_{\text{r}}\text{ ,}\label{f2}
	\end{equation}
	where $H={\dot{a}}/{a}$ is Hubble expansion rate. Here a dot
	denote derivative with respect to time. 
	
From Eqs. (\ref{f1})
	and (\ref{f2}), we identify the effective density and
	pressure of the Galileon field 
	\begin{equation}
		\rho_{x}=-\frac{1}{2}c_{2}\dot{\pi}^{2}+\frac{3}{M^{3}}c_{3}H\dot{\pi}^{3},\label{rho}
	\end{equation}
	\begin{equation}
		P_{x}=-\frac{1}{2}c_{2}\dot{\pi}^{2}-\frac{1}{M^{3}}c_{3}\dot{\pi}^{2}\ddot{\pi}.\label{pressure}
	\end{equation}
	In
	the FLRW space-time Eq.(\ref{eq:interaction}) reads
	\begin{equation}
		\dot{\rho}_{x}+3H\left(\rho_{x}+P_{x}\right)=Q,\label{pi}
	\end{equation}
	\begin{equation}
		\dot{\rho}_{\text{c}}+3H\rho_{\text{c}}=-Q.\label{dm}
	\end{equation}
Additionally,
	the conservation laws for radiation and baryon components on the
	background read 
	\begin{equation}
		\dot{\rho}_{\text{r}}+4H\rho_{\text{r}}=0,\quad \dot{\rho}_{\text{b}}+3 H\rho_{\text{b}}=0.\label{r}
	\end{equation}
	
Once a form of the interaction $Q$ is known, the background dynamics is fully determined
	by the energy conservation equations ($\ref{pi}$) and ($\ref{dm}$)
	and the Friedmann equations (\ref{f1}) and (\ref{f2}).
	
Let us take benefit of the existence of a de Sitter
	(dS) background characterized by $H=H_{\text{dS}}\equiv cst$, $\dot{\pi}=\dot{\pi}_{\text{dS}}\equiv cst$,
	and fix the free parameters $c_{2},\:c_{3}$ in the Galileon Lagrangian
	in terms of the interaction function. Writing the dynamical equations in the dS era, and solving the resulting
	equations we obtain 
	\begin{equation}
		x_{\text{dS}}^{2}c_{2}=6,\quad x_{\text{dS}}^{3}c_{3}=2+
		\frac{Q_{\text{dS}}}{9H_{\text{dS}}^{3}M_{\text{Pl}}^{3}},\label{c2}
	\end{equation}
	
	\begin{equation}
		\Omega_{\text{b},\text{dS}}=0,\quad\Omega_{\text{c,dS}}=-\frac{Q_{\text{dS}}}{9H_{\text{dS}}^{3}M_{\text{Pl}}^{2}}.\label{eq:Om_dS}
	\end{equation}
	where $x_{\text{dS}}=\frac{\dot{\pi}_{\text{dS}}}{H_{\text{dS}}M_{\text{Pl}}}$
	and we normalized $M$ to $M^{3}=M_{\text{Pl}}H_{\text{dS}}^{2}$
	\cite{de2012conditions}. These equations indicate that $Q_{\text{dS}}$ must be negative, implying an energy flow from dark energy (DE) to dark matter (DM) during the de Sitter (dS) era. Notably, the only pertinent free parameters are the coupling parameters within the interaction function, and the DM density during the dS era is non-zero, contingent upon the coupling function.
	
	Indeed, deriving the exact form of $Q$ from fundamental principles is elusive, and the existing forms are typically derived phenomenologically or inspired by investigations in scalar-tensor theories \cite{Holden_2000}. Among the diverse interaction terms explored in literature, we adopt the coupling function described in \cite{valiviita2008large}:
	\begin{equation}
		Q=H\alpha\rho_{x},\label{QQ}
	\end{equation}
	where $\alpha$ is the coupling constant and $\alpha H$ is the rate
	of transfer of DE to DM.
	
	In this case the Galileon field equation, Eq.(\ref{pi}) reads
	\begin{equation}
		2\ddot {\pi}\left (8 c_ 3 \
		{H} \dot {\pi}^2 -  c_ 2 M^3 \dot {\pi} \right)+6 c_ 3\dot {\pi}^3\left (\dot {H} - 
		(\alpha - 1) {H}^2 \right) + (\alpha - 
		2) c_ 2 M^3 H\dot {\pi}^2 =0, \label{eq_galileon}
	\end{equation}
	and  Eqs.(\ref{c2}) and (\ref{eq:Om_dS})
	reduce to
	
	\begin{equation}
		x_{\text{dS}}^{2}c_{2}=6,\quad x_{\text{dS}}^{3}c_{3}=\frac{\alpha-6}{\alpha-3},\quad\Omega_{\text{b},\text{dS}}=0,\quad\Omega_{\text{c,dS}}=\frac{\alpha}{\alpha-3}.\label{c2-1}
	\end{equation}
We proceed to examine the background dynamics employing autonomous dynamical systems. Initially, we simplify the evolution equations ($\ref{f1}$) and ($\ref{f2}$) into more manageable first-order differential equations by introducing new dimensionless dynamical variables. To accomplish this, we adopt the methodology outlined in the non-interacting case in \cite{nesseris2010observational}, introducing the dimensionless variables $r_{1}$ and $r_{2}$: 
	\begin{equation}
		r_{1}=\frac{\dot{\pi}_{\text{dS}}H_{\text{dS}}}{\dot{\pi}H},\quad r_{2}=\frac{1}{r_{1}}\left(\frac{\dot{\pi}}{\dot{\pi}_{\text{dS}}}\right)^{4},\label{variables}
	\end{equation}
and the definitions
	\begin{equation}
		\epsilon_H=\frac{\dot{H}}{H^2},\qquad \epsilon_{\pi}=\frac{\ddot{\pi}}{\dot{\pi}H}.
	\end{equation}
Then we easily obtain: 
	\begin{equation}
		\frac{H}{H_{\text{dS}}}=\frac{1}{r_{1}\left(r_{1}r_{2}\right)^{1/4}},\quad\frac{\dot{\pi}}{\dot{\pi}_{ds}}=\left(r_{1}r_{2}\right)^{1/4}.\label{eq:Hub_pi}
	\end{equation}
In the dS phase, where $H=H_{\text{dS}}\equiv \text{cst}$, $\dot{\pi}=\dot{\pi}_{\text{dS}}\equiv cst$,
	we get $r_{1}= r_{2}=1.$
	
The dimensionless variables $r_1,\,r_2,\,\Omega_b,\,\Omega_r$  obey
the differential equations
	\begin{flalign}
		r'_{1}= & -r_{1}\left(\epsilon_{\pi}+\epsilon_H\right),\label{eq:ode1}\\
		r'_{2}= & r_{2}\left(5\epsilon_{\pi}+\epsilon_H\right),\label{eq:ode2}\\
		\Omega'_{\text{r}}= & -\Omega_{r}\left(4+2\epsilon_H\right),\label{eq:ode3}\\
		\Omega'_{\text{b}}= & -\Omega_{b}\left(3+2\epsilon_H\right),\label{eq:ode4}
	\end{flalign}
where the density parameters  are 
	$
	\Omega_{a}={\rho_{\text{a}}}/\left({3M_{\textrm{Pl}}^{2}H^{2}}\right),\,\text{a}=\{r,\,,b\,,c\,,x\}$, and the prime denotes derivative with respect to $N=\ln a$.
	Combining (\ref{eq:ode1}) and (\ref{eq:ode2}) we obtain:
	\begin{equation}
		\epsilon_H=-\frac{5r_{1}'}{4r_{1}}-\frac{r_{2}'}{4r_{2}}.\label{eq:Hubble}
	\end{equation}
In the non-interacting case ($\alpha=0$), the standard scenario is restored, consistent with the analysis presented  in \cite{nesseris2010observational}. Next, we express \eqref{f1}, \eqref{f2}, and \eqref{eq_galileon} in terms of the dimensionless variables and proceed to solve for $\epsilon_H$ and $\epsilon_{\pi}$:
	\begin{flalign}
		\epsilon_H= & \left(\alpha-3\right)\left[\frac{r_{1}^{2}r_{2}\left(r_{1}-1\right)\left(\alpha-6\right)^{2}+2\left(3r_{1}^{3}r_{2}-\Omega_{r}-3\right)\left(\alpha r_{1}-\alpha-3r_{1}+6\right)}{4\left(\alpha-3\right)\left(\alpha\left(r_{1}-1\right)-3r_{1}+6\right)-r_{1}^{2}r_{2}\left(\alpha-6\right)^{2}}\right]\label{epsilonH},
	\end{flalign}
	\begin{flalign}
		\epsilon_{\pi}= & \left(\alpha-3\right)\left(\alpha-6\right)\left[\frac{3r_{1}^{3}r_{2}-\Omega_{r}-3+2\left(\alpha-3\right)\left(r_{1}-1\right)}{4\left(\alpha-3\right)\left(\alpha\left(r_{1}-1\right)-3r_{1}+6\right)-r_{1}^{2}r_{2}\left(\alpha-6\right)^{2}}\right]\label{epsilonpi}.
	\end{flalign}
The DE density $\Omega_{{x}}$ and the DE EoS $\omega_{{x}}$ are then given in terms of $r_{1}$ and $r_{2}$ as: 
	\begin{flalign}
		\Omega_{\text{x}} & =-\frac{r_{1}^{3}r_{2}\left[\left(\alpha-3\right)r_{1}+6-\alpha\right]}{\alpha-3},\label{Omega_Dark}\\
		\omega_{x} & =\frac{1+\left(\frac{\alpha-6}{\alpha-3}\right)\frac{\epsilon_{\pi}}{3  r_1}}{1-\left(\frac{\alpha-6}{\alpha-3}\right)\frac{1}{r_1}}.
	\end{flalign}
	Let us denote the autonomous system (\ref{eq:ode1})-(\ref{eq:ode4}) generically as:
	\begin{flalign}
		\dot{\vec{x}}_{i}& =f\left(\overrightarrow{x}\right),\quad\overrightarrow{x}=\left(r_{1},r_{2},\Omega_{r},\Omega_{b}\right).\label{gen_dyn}
	\end{flalign}
	Firstly, we identify the fixed or critical points of equations (\ref{gen_dyn}) and investigate their stability throughout cosmic history. The fixed points correspond to the roots of  $\dot{\vec{x}}_{i}=0$. The stability analysis is conducted using first-order perturbation technique around these fixed points, followed by the formation of the coefficient matrix for the perturbed terms. A fixed point is deemed stable (an attractor) if all eigenvalues of the perturbation matrix are negative, a saddle if the eigenvalues have mixed signs, and unstable if all eigenvalues are positive \cite{perko2012differential}. We have identified five fixed points, denoted as $A$, $B$, $C$, $D$, and $E$. Their properties are summarized in Tab. \ref{Tab1}, where $q=-1-\epsilon_{H}$ represents the deceleration parameter, and the eigenvalues $C_{\pm}$ are given by:
	\begin{equation}
		C_{\pm}=\frac{3\left(\alpha-3\right)}{2}\left[\frac{3\alpha-12\alpha+48\pm\alpha\sqrt{\left(3\alpha-20\right)\left(\alpha-12\right)}}{\alpha^{2}-24\alpha+72}\right].
	\end{equation}
	
	\begin{table}[h]
		\centering 
		\resizebox{14cm}{!}{
		\begin{tabular}{c|c|c|c|c}
			\hline \hline
			Fixed Points  & $\left(r_{1},\:r_{2}\:,\Omega_{r}\:,\Omega_{b}\right)$  & Eigenvalues  & Stability  & q\tabularnewline
			\hline 
			A  & $\left(0,0,1,0\right)$  & $(1,1,\frac{5-\alpha}{2},-\frac{9-5\alpha}{2})$  & $\begin{array}{c}
				\text{ Unst. }9/5<\alpha<5\\
				\text{Saddle otherwise}
			\end{array}$  & $1$\tabularnewline
			\hline 
			B  & $\left(0,0,0,\Omega_{b}\right)$  & $(0,1,\frac{9-2\alpha}{4},\frac{10\alpha-21}{4})$  & $\begin{array}{c}
				\text{ Unst. }21/10<\alpha<9/2\\
				\text{Saddle otherwise}
			\end{array}$  & $1/2$\tabularnewline
			\hline 
			C  & $\left(\frac{\alpha^{2}-11\alpha+30}{\alpha^{2}-13\alpha+30},0,1,0\right)$  & $(8,-\frac{\alpha^{2}}{10}+\frac{3\alpha}{2}-5,1,1)$  & $\begin{array}{c}
				\text{ Unst. if }5 < \alpha < 10\\
				\text{Saddle otherwise}
			\end{array}$  & $1$\tabularnewline
			\hline 
			D  & $\left(\frac{2\alpha^{2}-21\alpha+54}{2\alpha^{2}-24\alpha+54},\:0,\:0,\:\Omega_{b}\right)$  & $(0,-1,6,-\frac{\alpha^{2}}{9}+\frac{3\alpha}{2}-\frac{9}{2})$  & Saddle  & $1$\tabularnewline
			\hline 
			E  & $\left(1,1,0,0\right)$  & $(-4,-3,\:C_{+},\:C_{-})$  & $\begin{array}{c}
				\text{ Stable if }\alpha<3\\
				\text{Unst. otherwise}
			\end{array}$  & $-1$\tabularnewline
			\hline \hline
		\end{tabular}
		}
		\caption{\label{Tab1} Location, Eigenvalues and Stability of the Critical
			Points.}
	\end{table}
	
	The two fixed points $A$ and $B$ represent the eras of radiation and matter domination, respectively, with no contribution from dark energy (DE). They are unstable for $9/5 <\alpha<5$ and $21/10 <\alpha<9/2$, respectively, and belong to the small regime \cite{de2012conditions,nesseris2010observational}. Fixed points $C$ and $D$ also correspond to radiation and matter dominated eras but include contributions from DE. They are unstable for $5<\alpha<10$ and saddle, respectively. The last fixed point, $E$, corresponds to the de Sitter (dS) fixed point. It is stable if $\alpha<3$ and potentially acts as an attractor for the entire cosmological evolution, regardless of the initial conditions.

We have two viable paths for a valid cosmological evolution. The first/second path begins from the unstable radiation-dominated era, $A$/$C$, transitions to the unstable matter-dominated era, $B$/$D$, and concludes at the dS point $E$. It's noteworthy that for $\alpha=0$, the dynamical analysis conducted above converges with that in \cite{nesseris2010observational}. In this case, The point $C$ coincides with $A$, and the point $D$ coincides with $B$, and become saddles, while the dS fixed point $E$ remains stable.
	
	\subsection{Analysis of the fixed points eras\label{sec:Analysis-of-the}}
	
	We'll embark on a detailed analysis of the dynamics within the eras delineated by the fixed points and deduce constraints on the DE-DM coupling.
	\begin{itemize}
		\item \textbf{\textit{Small regime}} 
	\end{itemize}
	In this regime, the fixed points A and B are characterized by $r_{1}\ll1$ and $r_{2}\ll1$. Under these conditions, the autonomous system of equations simplifies to:
	\begin{flalign}
		r_{1}'= & -\frac{r_{1}}{4}\left(2\alpha-\Omega_{r}-9\right),\label{eq:small-r1}\\
		r_{2}'= & \frac{r_{2}}{4}\left(10\alpha+3\Omega_{r}-21\right),\label{eq:small_r2}\\
		\Omega_{\text{r}}'= & \Omega_{\text{r}}\left(\Omega_{\text{r}}-1\right),\\
		\Omega_{\text{b}}'= & \Omega_{\text{b}}\Omega_{\text{r}}.
	\end{flalign}
	These equations integrate to:
\begin{eqnarray}
\Omega_{\text{r}} &=&\frac{1}{1+d_{4}a}, \quad \Omega_{\text{b}} = \frac{d_{1}a}{1+d_{4}a}, \label{eq:small1}\\
r_{1} &=&\frac{d_{3}a^{\frac{5-\alpha}{2}}}{(1+d_{4}a)^{1/4}}, \quad r_{2} = \frac{d_{2}a^{\frac{5\alpha-9}{2}}}{(1+d_{4}a)^{3/4}}, \label{eq:small2}
\end{eqnarray}
where $d_{i}$ are constants. The two fixed points in the small regime, A (unstable for $\alpha<5$) and B (unstable for $\alpha<9/2$), represent pure radiation-dominated and pure matter-dominated solutions, respectively. In the vicinity of A, we set $d_{4}\approx0$ and obtain: 
	\begin{flalign}
		r_{1}\approx & a^{\frac{5-\alpha}{2}},\quad r_{2}\approx a^{\frac{5\alpha-9}{2}},\quad H\approx a^{-2},\quad \pi\approx a^{\frac{\alpha+3}{4}},\label{eq:small2-2}
	\end{flalign}
	While for B, we consider $d_{4}$ to be very large but with $d_{1}/d_{4}\sim1$, yielding:
	
	\begin{flalign}
		r_{1}\approx  a^{\frac{9-2\alpha}{4}},\quad r_{2}\approx a^{\frac{10\alpha-21}{4}},\quad H\approx a^{-3/2},\quad \pi\approx a^{\frac{2\alpha+3}{4}}.\label{eq:small2-2-1}
	\end{flalign}
	We observe that the exponents of the scale factor in (\ref{eq:small2-2}) and (\ref{eq:small2-2-1}) precisely match the eigenvalues of fixed points A and B, respectively. Additionally, for $9/5 < \alpha < 5$, $r_{2}$ grows faster than $r_{1}$. We also derive the conventional evolution laws of the Hubble parameter during the radiation and matter domination epochs, regardless of the DE-DM coupling constant. This outcome is anticipated since the interaction term, proportional to $\rho_{DE}$, only becomes significant at the onset of the DE-dominated epoch.

In the small regime, the DE and the total effective equation of state (EoS) are approximated by:
	\begin{flalign}
		w_{{x}}\approx & -\frac{1}{12}\left[\frac{2\left(r_{1}^{2}+1\right)\alpha^{2}+\left(2\left(\Omega_{r}+6\right)r_{1}+\Omega_{r}-15\right)\alpha-6\Omega_{r}\left(r_{1}+1\right)+18\left(1-3r_{1}\right)}{\alpha-6}\right],\\
		w_{\text{eff}}\approx & \frac{\Omega_{r}}{3}.
	\end{flalign}
	For the radiation dominated fixed point A we have:
	\begin{flalign}
		w_{{x}}=\frac{1}{6}-\frac{\alpha}{6},\quad w_{\text{eff}}=\frac{1}{3},
	\end{flalign}
	while for the matter dominated fixed point B we have:
	\begin{flalign}
		w_{\text{x}}=\frac{1}{4}-\frac{\alpha}{6},\quad w_{\text{eff}}=0 & .\label{eq:small_speed_2}
	\end{flalign}
		In the non-interacting cubic Galileon model ($\alpha=0$), we approximate the solutions as follows: $r_{1}\approx a^{5/2}$, $r_{2}\approx a^{-9/2}$, $H\approx a^{-2}$, $w_{\text{x}}=1/6$ for point A, and $r_{1}\approx a^{9/4}$, $r_{2}\approx a^{-21/4}$, $H\approx a^{-3/2}$, $w_{\text{x}}=1/4$ for point B. These behaviors differ from those in ref. \cite{de2012conditions} where $r_{1}\approx a^{5/4}$, $r_{2}\approx a^{7/4}$, $H\approx a^{-2}$, and $w_{\text{x}}=1/4$ in the radiation era, and $r_{1}\approx a^{9/4}$, $r_{2}\approx a^{3/8}$, $H\approx a^{-93/32}$, and $w_{\text{x}}=1/8$ in the matter era. The discrepancy originates from numerical factors ($1/8$ instead of $1/4$) in Eqs. (\ref{eq:small-r1}) and (\ref{eq:small_r2}).
	\begin{itemize}
		\item \textbf{\textit{Fixed Point C}} 
	\end{itemize}
	This point exhibits instability within the interval $5<\alpha<10$, corresponding to a pure radiation-dominated solution. In its vicinity, the parameter $r_{2}$ becomes small. Consequently, the DE density, DE EoS, and the  effective EoS are approximately described by:
	\begin{flalign}
		\Omega_{{x}} & =5\left[\frac{\left(\alpha-6\right)^{3}\left(\alpha-5\right)^{2}}{\left(\alpha-3\right)^{3}\left(\alpha-10\right)^{3}}\right]r_{2},\\
		w_{{x}} & =\frac{\alpha}{3}-\frac{7}{3}-\frac{1}{60}\left[\frac{\left(\alpha-6\right)^{3}\left(\alpha-5\right)^{2}\left(\alpha-7\right)}{\left(\alpha-3\right)^{3}\left(\alpha-10\right)}\right]r_{2},\\
		w_{{eff}} & =\frac{1}{3}+\frac{5}{3}\left[\frac{\left(\alpha-6\right)^{3}\left(\alpha-5\right)^{2}\left(\alpha-7\right)}{\left(\alpha-3\right)^{3}\left(\alpha-10\right)^{3}}\right]r_{2}.
	\end{flalign}
	At the fixed point, $r_{2}=0$, we are left with the results 
	\begin{equation}
		\Omega_{{x}}=0,\quad w_{{x}}=-\frac{7}{3}+\frac{\alpha}{3},\quad w_{\text{eff}}=\frac{1}{3}.
	\end{equation}
	
	\begin{itemize}
		\item \textbf{\textit{Fixed Point D}} 
	\end{itemize}
	This fixed point is saddle and corresponds to a pure matter-dominated solution. In this phase, we can expand the DE density, the DE EoS, and the effective EoS  around $r_{2}\ll1$ to obtain:
	\begin{flalign}
		\Omega_{{x}} & =\frac{9}{8}\left[\frac{\left(2\alpha-9\right)^{2}\left(\alpha-6\right)^{3}}{\left(\alpha-3\right)^{3}\left(\alpha-9\right)^{3}}\right]r_{2},\\
		w_{{x}} & =\frac{\alpha}{3}-2-\frac{1}{216}\left[\frac{\left(2\alpha-9\right)^{2}\left(\alpha-6\right)^{4}}{\left(\alpha-3\right)^{3}\left(\alpha-9\right)}\right]r_{2},\\
		w_{\text{eff}} & =\frac{3}{8}\left[\frac{\left(2\alpha-9\right)^{2}\left(\alpha-6\right)^{3}}{\left(\alpha-3\right)^{3}\left(\alpha-9\right)^{3}}\right]r_{2}.
	\end{flalign}
	At exactly the fixed point, $r_{2}=0$, we have
	\begin{equation}
		\Omega_{{x}}=0,\quad w_{{x}}=-2+\frac{\alpha}{3},\quad w_{\text{eff}}=0.
	\end{equation}
	
	\begin{itemize}
		\item \textbf{\textit{dS Fixed Point}} 
	\end{itemize}
	The dS fixed point characterized by $r_{1}=1,\:r_{2}=1$ is stable
	for $\alpha<3$. At this point we have
	
	\begin{flalign}
		\Omega_{\text{c,dS}}=\frac{\alpha}{\alpha-3},\quad\Omega_{x,\text{ds}}=\frac{3}{3-\alpha},\quad w_{x,\text{ds}} & =-1+\frac{\alpha}{3},\quad w_{\text{\text{e}ff,dS}}=-1.
	\end{flalign}
	We detect deviations from the standard $\Lambda$CDM model when $\alpha\neq 0$. Specifically, during the de Sitter (dS) era, DE  continues to dominate, albeit with a minor contribution from DM. This arises from the gradual decay of the DE fields into DM, resulting in a constant fraction $\Omega_{\text{c,dS}}$ throughout the dS era. Additionally, we observe that the EoS $w_{x}$ deviates slightly from  $-1$. By combining constraints on the solutions and ensuring the positivity of the DM density during the dS era, we infer that $\alpha < 0$.

	\begin{itemize}
		\item \textbf{\textit{Tracker Solution}} 
	\end{itemize}
	Assuming a small coupling $\alpha$ in the dark sector, we note from Table (\ref{Tab1}) that the coordinate $r_{1}$ for the fixed points C and D is of order unity. Following the approach outlined in \cite{nesseris2010observational}, we approximate the dynamics of the model by setting $r_{1}=1$ in the dynamical equations and solving for $r_{2}$, $\Omega_{r}$, and $\Omega_{b}$. In this scenario, the autonomous system of equations (\ref{eq:ode2})-(\ref{eq:ode4}) can be expressed as:
	\begin{flalign}
		\frac{r_{2}'}{r_{2}}= & \left(\alpha-3\right)\left[\frac{(5\alpha-24)\left(\Omega_{r}-3r_{2}+3\right)}{\alpha^{2}r_{2}-12\left(\alpha-3\right)\left(r_{2}+1\right)}\right],\label{eq:T1}\\
		\frac{\Omega_{\text{r}}'}{\Omega_{\text{r}}}= & -4\left[\frac{\alpha^{2}r_{2}+3\left(\alpha-3\right)\left(\Omega_{r}-7r_{2}-1\right)}{\alpha^{2}r_{2}-12\left(\alpha-3\right)\left(r_{2}+1\right)}\right],\label{eq:T2}\\
		\frac{\Omega_{\text{b}}'}{\Omega_{\text{b}}}= & -3\left[\frac{\alpha^{2}r_{2}+4\left(\alpha-3\right)\left(\Omega_{r}-6r_{2}\right)}{\alpha^{2}r_{2}-12\left(\alpha-3\right)\left(r_{2}+1\right)}\right].\label{eq:T3}
	\end{flalign}
	Combining these equations we obtain 
	\begin{equation}
		\frac{r_{2}'}{r_{2}}-\gamma\left(\frac{\Omega_{r}'}{\Omega_{r}}-4\right)=0,\,\, \frac{r_{2}'}{r_{2}}-\gamma\left(\frac{\Omega_{b}'}{\Omega_{b}}+3\right)=0,\,\, \gamma=2-\frac{5\alpha}{12}.\label{eq:r2/Omr}
	\end{equation}
	The solution of Eqs.(\ref{eq:r2/Omr}) are
	\begin{equation}
		r_{2}\left(a\right)=r_{2}\left(0\right)a^{4\gamma}\left(\frac{\Omega_{r}\left(a\right)}{\Omega_{r}\left(0\right)}\right)^{\gamma},
		\,\, r_{2}\left(a\right)=r_{2}\left(0\right)a^{3\gamma }\left(\frac{\Omega_{b}\left(a\right)}{\Omega_{b}\left(0\right)}\right)^{\gamma}.\label{eq:r2Sol}
	\end{equation}
	Using (\ref{eq:Hubble}), we obtain:
	\begin{equation}
		H(a)=H(0)a^{-\gamma }\left(\frac{\Omega_{r}\left(a\right)}{\Omega_{r}\left(0\right)}\right)^{-\gamma/4}.
	\end{equation}
	The DE density, the DE EoS, and the effective EoS parameters
	along the tracker are now given by 
	\begin{flalign}
		\Omega_{{x}}= & \frac{3r_{2}}{3-\alpha},\\
		w_{{x}}= & \left(\frac{\alpha-3}{9}\right)\left[\frac{\left(\Omega_{r}+3\right)\alpha^{2}-12\left(\alpha-3\right)\left(\Omega_{r}+6\right)}{\left[r_{2}\alpha^{2}-12\left(\alpha-3\right)\left(r_{2}+1\right)\right]}\right],\nonumber \\
		w_{\text{eff}}= & -\left[\frac{r_{2}\alpha^{2}+4\left(\alpha-3\right)\left(\Omega_{r}-6r_{2}\right)}{r_{2}\alpha^{2}-12\left(\alpha-3\right)\left(1+r_{2}\right)}\right].
	\end{flalign}
	We observe that the DE density reach the solution in the
	dS era for $r_{2}=1$. When $\alpha=0$, we reproduce exactly the relations of \cite{nesseris2010observational}
	\begin{flalign}
		\text{\ensuremath{\Omega_{x}}}= & r_{2},\quad w_{{x}}=-\frac{\Omega_{r}+6}{3\left(r_{2}+1\right)},\quad w_{\text{eff}}=\frac{\left(\Omega_{r}-6r_{2}\right)}{3\left(1+r_{2}\right)}.
	\end{flalign}
	
Although we can express all the relevant quantities along the tracker in terms of $\Omega_r$, obtaining an algebraic equation for $\Omega_r$ and solving it as done in \cite{nesseris2010observational}  is not feasible. Therefore, we substitute (\ref{eq:r2Sol}) into (\ref{eq:T2}) and (\ref{eq:T3}) and numerically integrate the resulting equations. Finally, assuming that $\Omega_{c\,\textrm{dS}}\geq 0$, we derive the constraint $\alpha \leq 0$.

	\section{ICG on the perturbed flat space-time\label{sec:Growth-rate-of}}
	
	The investigation of the growth rate of cosmological density perturbations has emerged as a potent tool for distinguishing between cosmological models based on modified theories of gravity and DE based models. While all models may perfectly mimic the $\Lambda$CDM evolution at the background level, they inherently influence structure formation. An important aspect in this regard is the evolution of the linear matter density contrast $\delta_{m}\equiv\delta\rho_{m}/\rho_{m}$, which satisfies the equation:
\begin{equation}
\ddot{\delta}_m+2H\dot{\delta}_m-4\pi G_{\text{eff}}\rho_{m}\approx0. \label{eq:delta}
\end{equation}
Here, $G_{\text{eff}}$ is the effective gravitational constant,  function of the scale factor and the cosmological scale. The matter density contrast is linked to the observed quantity $f(a)\sigma_{8}(a)$, where $f(a)=d\ln\delta(a)/d\ln(a)$ and $\sigma_{8}(a)=\sigma_{8}\delta_{m}(a)/\delta_{m}(1)$ represents the root mean square fluctuations of the linear density field within a radius of $8h^{-1}\text{Mpc}$, with $\sigma_{8}$ being its present value.

Scalar perturbations on a flat FLRW spacetime in the conformal Newtonian gauge are described by the following metric:
\begin{equation}
ds^{2}=-a^{2}(\eta)\left[(1+2\psi)d\eta^{2}-\delta_{ij}(1-2\theta)dx^{i}dx^{j}\right]. \label{per ds}
\end{equation}
Here, $\psi$ and $\theta$ describe scalar perturbations, with $\psi$ being the gravitational potential.

In cosmological perturbations theory, each quantity, denoted by $\chi$, is expanded up to the desired order with a homogeneous background $\overline{\chi}$ and a small part $\delta \chi=\sum_{n=1}^{\infty}\frac{\chi _n}{n!}$, where $n$ denotes the perturbation order. Here, we specifically focus on linear perturbation theory, i.e., where $n=1$.
\subsection{Perturbed energy momentum tensor}
		We are concerned with matter and dark energy dominated eras where the anisotropic stress can be neglected. In such cases, the energy-momentum tensor of a fluid $A$ is given by:
	\begin{eqnarray}
	T_{\mu \nu }^{(A)}& =&T_{\mu \nu} ^{\left( A\right) } 
		=\left( \rho_A +P_A\right) u_{\mu }u_{\nu }+P_A g_{\mu \nu },\\
		\rho _{A} &=&{\overline{\rho}}_{A}+\delta \rho _{A},\,\,
		P_{A} =\overline{P}_{A}+\delta P_{A}.
	\end{eqnarray}%
	Here $\overline{\rho}_A$ , $\overline{P}_A$ are energy density and pressure on the background,  $u_{A}^{\mu }$ is the $A-$fluid
	four velocity given by:
	\begin{equation}
		u_{A}^{\mu } =a^{-1}\left( 1-\psi ,\upsilon _{A}^{i}\right),\,\, 
		\upsilon _{A}^{i}=\partial ^{i}u_{A},\text{ } \label{u}
	\end{equation}%
	and $u_{A}$ is the peculiar velocity potential.
	
	The total energy-momentum tensor is the sum of the energy-momentum tensor of
	the individual fluids:
	\begin{equation}
		T_{\mu \nu } =\sum\limits_{A}T_{\mu \nu} ^{\left( A\right) } 
		=\left( \rho +P\right) u_{\mu }u_{\nu }+Pg_{\mu \nu },
	\end{equation}%
	where%
	\begin{equation}
		\rho =\sum\limits_{A}\rho_{A},\qquad P=\sum\limits_{A} P_{A},\qquad u^{\mu}=a^{-1}\left( 1-\psi ,\partial ^{i}u\right),
	\end{equation}
	and the total velocity potential $u$ is defined by:
	\begin{equation}
		\left( \rho +p\right) u=\sum \left( \rho _{A}+p_{A}\right) u_{A}.
	\end{equation}%
		The energy-momentum tensor of each component is not conserved, and its divergence introduces a local energy-momentum transfer tensor $Q_{A}^{\mu}$:
\begin{equation}
\nabla_{\nu}T_{A}^{\mu\nu}=Q_{A}^{\mu}. \label{cons_T}
\end{equation}
Here, $Q_{A}^{\mu}$ is referred to as the covariant interaction function, often expressed as:
\begin{equation}
Q_{A}^{\mu}=Q_{A}u^{\mu}+f_{A}^{\mu}, \label{per Q}
\end{equation}
where the functions $Q_A$ and $f_A^{\mu}$ denote the energy density transfer rate and the momentum density transfer rate to the $A$-component as observed in the center of mass frame:
\begin{equation}
Q_{A}=\bar{Q}_{A}+\delta Q_{A},\,\,f_{A}^{\mu}=a^{-1}\left( f_{A}^{0},\partial^{i}f_{A}\right),\,\,u_{\mu}f_{A}^{\mu}=0,\label{Q}
\end{equation}
where, $f_{A}$ represents a momentum transfer potential. 
	From the conservation law of the total energy-momentum tensor $\nabla _{\nu
	}T_{\text{ \ }}^{\mu \nu }=0$, the  coupling function $Q_{\text{
			\ }A}^{\mu }$ is constrained by:
	\begin{equation}
		\sum\limits_{A}Q_{{A}}=\sum\limits_{A}\delta
		Q_{A}=\sum\limits_{A}f_{A}^{\text{ }}=0.
	\end{equation}%
	Given the orthogonality relation between $f$ $_{A}^{\mu }$ and   $u_{\mu
	}$, it follows that $f_{A\text{ }}^{\text{ }0\text{ }}=0,$
	and hence:
	\begin{equation}
		f\text{ }_{A}^{\mu }=a^{-1}\left( 0,\partial ^{i}f_{A}\right).  \label{f}
	\end{equation}%
	By combining  (\ref{Q}) and (\ref{f}) with (\ref{u}), we arrive at:
\begin{equation}
		Q_{A}^{0}=a^{-1}\left[ \bar{Q}_{A}\left( 1-\psi \right) +\delta
		Q_{A\text{ }}^{\text{ \ }}\right],  \label{Q0}
	\end{equation}%
	\begin{equation}
		Q_{A}^{i}=a^{-1}\partial ^{i}\left( \bar{Q}_{A}u+f_{A}^{\text{ }%
		}\right).  \label{Qi}
\end{equation}%
	At zero order we have $Q_{
		A}^{\mu }=a^{-1}\left( \bar{Q}_{A},\vec{0}\right)$. This indicates the absence of momentum transfer on the background.
		
For coupled  DM and DE fluids, $A={x},c$,  we have:
	\begin{equation}
		\nabla _{\mu }T^{\mu \nu}_{\left( x\right) } =Q^{%
			\nu}_{\left( x\right) }, \quad \nabla _{\mu }T^{\mu \nu}_{\left( c\right) } = Q^{%
			\nu}_{\left( c\right) }, \label{per cons de}
	\end{equation}%
	where $ Q^{
		\nu }_{\left( c\right) }=-Q^{\nu }_{\left( x\right) }.$
On the background Eq.(\ref{per cons de}) becomes:
	\begin{eqnarray}
		\bar{\rho}_{c}^{\prime }+3\mathcal{H}\bar{\rho}_{c}&=&-a\bar{Q}_{x},  \label{perdm} \\
		\bar{\rho}_{x}^{\prime }+3\mathcal{H}\left(\bar{\rho}_{x}+\bar{P}_{x}\right)&=& a\bar{Q}_{x},\label{per de}
	\end{eqnarray}%
	where $\mathcal{H}= a H $ is Hubble parameter in terms of conformal time. Le us define the density contrast $\delta_{A}=\delta\rho_{A}/\rho_{A}$ of the fluid A, with 
	$\delta T_{0 A }^{0} =-\delta \rho _{%
		A}, 
	\delta T_{i A }^{0} =\left( \bar{\rho}_{%
		A}+\bar{P}_{A}\right) \upsilon _{i A}=-\delta T_{ 0  A }^{i}$ and 
$\delta T^{i }_{j A} =\delta P_{A}\delta _{j}^{i}$. Then, the perturbed energy and momentum balance equations of DM are obtained from Eq.(\ref{per cons de}) as follows:
\begin{equation}
\delta _{c}^{\prime }+\nabla^{2}u_{c}-3\theta ^{\prime }=\frac{a}{\bar{\rho}_c}\left(\bar{Q}_{c}\psi +\delta Q_{c}-\bar{Q}_{c}\delta_{c}\right) \label{per_delta_dm}
\end{equation}
\begin{equation}
		\left( \bar{\rho}_{c}u_{c}\right) ^{^{\prime }}+\psi \bar{\rho}_{c}+4%
		\mathcal{H}\bar{\rho}_{c}u_{c}=a\bar{Q}_{c}u+af_{c}.  \label{nu=i}
	\end{equation}%
Substituting Eq.(\ref{perdm}) into Eq.(\ref{nu=i}), we arrive at:
\begin{equation}
		u_{c}^{^{\prime }}+\mathcal{H}u_{c}+\psi =\frac{a}{\bar{\rho}_{c}}\left( 
		\bar{Q}_{c}\left( u-u_{c}\right) +f_{c}\right). 
		\text{ \ \ \ \ }  \label{per_dmi}
	\end{equation}%
Now, turning our attention to the DE component, let $\Pi$ represent the perturbed Galileon field, $	
		\Pi \left( \eta ,x^{i}\right) =\pi \left( \eta \right) +\delta \pi \left(
		\eta ,x^{i}\right)$, 	and  $
		T_{\alpha \beta }^{({x}) }=\bar{T}_{\alpha \beta }^{\left( x
			\right) }+\delta T_{\alpha \beta }^{(x) } 
	$, 
its energy momentum tensor, where  $\bar{T}_{00}^{({x}) } = \rho _{x }$, 
	$\bar{T}_{0i}^{\left(x \right) } = 0 ={T}_{i0}^{(x) }$ and $\bar{T}_{ij}^{({x}) } = P_{x } g_{ij} $
	 are given by Eqs.(\ref{rho}-\ref{pressure}). 
\noindent	 
The perturbed part of  $T_{\alpha \beta }^{(x)}$ read as:
\begin{eqnarray}
\delta T_{00}^{({x}) } &=&-\pi ^{\prime }\left[ \left( c_{2}-%
		\frac{9c_{3}}{a^{2}M^{3}}\mathcal{H}\pi ^{\prime }\right) \delta \pi
		^{\prime }+\frac{c_{3}}{a^{2}M^{3}}\left(\pi ^{\prime }\nabla ^{2}\delta \pi +%
		6\mathcal{H}\pi ^{\prime }\psi +\pi ^{\prime 2}\theta ^{\prime }\right)\right],  \label{delta_rho_de} \\
		\delta T_{0i}^{({x}) } &=&-\pi ^{\prime }\partial _{i}\left[ 
		\frac{c_{3}}{a^{2}M^{3}}\pi ^{\prime }\delta \pi ^{\prime }+\left( c_{2}-%
		\frac{3c_{3}}{a^{2}M^{3}}\mathcal{H}\pi ^{\prime }\right) \delta \pi -\frac{%
			c_{3}}{a^{2}M^{3}}\pi ^{2\prime }\psi \right], \\
		\delta T_{ij}^{({x}) } &=&\pi ^{\prime }\left[ -\frac{c_{3}}{%
			a^{2}M^{3}}\pi ^{\prime }\delta \pi ^{\prime \prime }+\frac{c_{3}}{a^{2}M^{3}}\pi ^{2\prime }\psi ^{\prime }+\left( -c_{2}+\frac{%
			c_{3}}{a^{2}M^{3}}\left( 3\mathcal{H}\pi ^{\prime }-2\pi ^{\prime \prime
		}\right) \right) \delta \pi ^{\prime }\right.\nonumber  \\
		&&\left. +\left( c_{2}+\frac{4c_{3}}{a^{2} M^{3}}
		\left( \pi ^{\prime \prime }-\mathcal{H}\pi ^{\prime }\right) \right) \pi
		^{\prime }\psi+\left(
		c_{2}+\frac{2c_{3}}{a^{2}M^{3}}\left( \pi ^{\prime \prime }-\mathcal{H}\pi
		^{\prime }\right) \right) \pi ^{\prime }\theta \right] \delta _{ij}.
	\end{eqnarray}%
\noindent
The perturbed  equation of motion of the Galileon field  follows from (\ref%
	{per cons de}) for $\nu =0$:
	\begin{eqnarray}
		&&\left( c_{2}-\frac{6c_{3}}{a^{2}M^{3}}\mathcal{H}\pi ^{\prime }\right) \pi
		^{\prime }\left( \delta \pi \right) ^{\prime \prime }+\left[ c_{2}\left( \pi
		^{\prime \prime }+4\mathcal{H}\pi ^{\prime }\right) -\frac{3c_{3}}{a^{2}M^{3}%
		}\left( 4\mathcal{H}\pi ^{^{\prime \prime }}+3\mathcal{H}^{\prime }\pi
		^{\prime }\right) \pi ^{\prime }\right] \left( \delta \pi \right) ^{\prime }+
		\nonumber \\
		&&\left[ -c_{2}+\frac{2c_{3}}{a^{2}M^{3}}\left( \pi ^{\prime \prime }+%
		\mathcal{H}\pi ^{\prime }\right) \right] \pi ^{\prime }\nabla ^{2}\left(
		\delta \pi \right) +\left( -c_{2}+\frac{9c_{3}}{a^{2}M^{3}}\mathcal{H}\pi
		^{\prime }\right) \pi ^{\prime 2}\psi ^{\prime }+  \nonumber \\
		&&\left[ c_{2}\left( -2\pi ^{\prime \prime }-4\mathcal{H}\pi ^{\prime
		}\right) +\frac{12c_{3}}{a^{2}M^{3}}\left( 2\mathcal{H}\pi ^{\prime }\pi
		^{^{\prime \prime }}+\mathcal{H}^{\prime }\pi ^{\prime 2}\right) \right] \pi
		^{\prime }\psi +\frac{c_{3}}{a^{2}M^{3}}\pi ^{\prime 3}\nabla ^{2}\psi + 
		\nonumber \\
		&&\frac{3c_{3}}{a^{2}M^{3}}\pi ^{\prime 3}\theta ^{\prime \prime }+3\left[
		-c_{2}+\frac{c_{3}}{a^{2}M^{3}}\left( 2\pi ^{^{\prime \prime }}+3\mathcal{H}%
		\pi ^{\prime }\right) \right] \pi ^{\prime 2}\theta ^{\prime }  \nonumber \\
		&=&-a^{3}\left( \bar{Q}_{\left( x\right) }\psi +Q_{\left( {x}\right) \text{ }%
		}^{\text{ \ }\left( 1\right) }\right). \label{motion_pert}
	\end{eqnarray}%
The component $\nu =i$ of (\ref{per cons de}) leads to:
	\begin{equation}
		\left[ c_{2}\left( \pi ^{\prime \prime }+2\mathcal{H}\pi ^{\prime }\right) +%
		\frac{c_{3}}{a^{2}M^{3}}\left( -3\mathcal{H}^{\prime }\pi ^{\prime 2}-3%
		\mathcal{H}^{2}\pi ^{\prime 2}-6\mathcal{H}\pi ^{\prime }\pi ^{^{\prime
				\prime }}+3\mathcal{H}^{2}\pi ^{^{\prime \prime }}\right) \right] \delta \pi
		=a^{3}\left( \bar{Q}_{x}u+f_{x}^{\text{ }\left( 1\right) }\right).\label{motion_pert_space}
	\end{equation}%
	Utilizing the Galileon field equation on the background: 
	\begin{equation}
		c_{2}\left( \pi ^{\prime \prime }+2\mathcal{H}\pi ^{\prime }\right) \pi
		^{\prime }-\frac{c_{3}}{a^{2}M^{3}}\left( 6\mathcal{H}\pi ^{^{\prime \prime
		}}+3\mathcal{H}^{\prime }\pi ^{\prime }\right) \pi ^{\prime 2}=-a^{3}\bar{Q}%
		_{\left( {x}\right) }, \label{eq ch non per}
	\end{equation}%
the equations (\ref{motion_pert}-\ref{motion_pert_space})  simplify to:
	\begin{eqnarray}
		&&\left( c_{2}-\frac{6c_{3}}{a^{2}M^{3}}\mathcal{H}\pi ^{\prime }\right) \pi
		^{\prime 2}\left( \delta \pi \right) ^{\prime \prime }+\left[ 2c_{2}\mathcal{%
			H}-\frac{3c_{3}}{a^{2}M^{3}}\left( 2\mathcal{H}\pi ^{^{\prime \prime }}+2%
		\mathcal{H}^{\prime }\pi ^{\prime }\right) \right] \pi ^{\prime 2}\left(
		\delta \pi \right) ^{\prime }+  \nonumber \\
		&&\left[ -c_{2}+\frac{2c_{3}}{a^{2}M^{3}}\left( \pi ^{\prime \prime }+%
		\mathcal{H}\pi ^{\prime }\right) \right] \pi ^{\prime 2}\nabla ^{2}\left(
		\delta \pi \right) +\left( -c_{2}+\frac{9c_{3}}{a^{2}M^{3}}\mathcal{H}\pi
		^{\prime }\right) \pi ^{\prime 3}\psi ^{\prime }+  \nonumber \\
		&&\left[ c_{2}\left( -2\pi ^{\prime \prime }-4\mathcal{H}\pi ^{\prime
		}\right) +\frac{12c_{3}}{a^{2}M^{3}}\left( 2\mathcal{H}\pi ^{\prime }\pi
		^{^{\prime \prime }}+\mathcal{H}^{\prime }\pi ^{\prime 2}\right) \right] \pi
		^{\prime 2}\psi +\frac{c_{3}}{a^{2}M^{3}}\pi ^{\prime 4}\nabla ^{2}\psi + 
		\nonumber \\
		&&\frac{3c_{3}}{a^{2}M^{3}}\pi ^{\prime 4}\theta ^{\prime \prime }-3\left[
		c_{2}-\frac{c_{3}}{a^{2}M^{3}}\left( 2\pi ^{^{\prime \prime }}+3\mathcal{H}%
		\pi ^{\prime }\right) \right] \pi ^{\prime 3}\theta ^{\prime }  \nonumber \\
		&=&-a^{3}\left( \left( \bar{Q}_{\left( {x}\right) }\psi +Q_{\left( {x}\right) 
			\text{ }}^{\text{ \ }\left( 1\right) }\right) \pi ^{\prime }-\bar{Q}_{\left(
			{x}\right) }\left( \delta \pi \right) ^{\prime }\right),  \label{per eq ch}
	\end{eqnarray}%
and
	\begin{equation}
		\left( \frac{3c_{3}\mathcal{H}^{2}}{a^{2}M^{3}}\left( -\pi ^{\prime 2}+\pi
		^{^{\prime \prime }}\right) \pi ^{\prime }+a^{3}\bar{Q}_{\left( {x}\right)
		}\right) \delta \pi =a^{3}\left( \bar{Q}_{x}u+f_{x}^{\text{ }\left(
			1\right) }\right) \pi ^{\prime }.  \label{i composante}
	\end{equation}%
In the absence of DE-DM coupling, we recover the equations in \cite{bartolo2013matter}. 
\noindent
Finally, the  equations of the 
	density and velocity perturbations for the baryon fluid remain the usual ones in the absence of DM-DE interaction:
	\begin{eqnarray}
		\delta _{b}^{\text{ }\prime }+\nabla ^{2}u_{b}-3\theta ^{\prime } &=&0,
		\label{delta b} \\
		u_{b}^{^{\prime }}+\mathcal{H}u_{b}+\psi &=&0. 
		 \label{u b}
	\end{eqnarray}
	
\subsection{Perturbed Einstein equations}
To linear order in the perturbation, the   $\left( 0,0\right)$  component of Einstein equations reads as: 
	\begin{eqnarray}
		&&2M_{\text{Pl}}^{2}\nabla ^{2}\theta -\left( 2a^{2}\bar{\rho}_{m}-\frac{6c_{3}}{%
			a^{2}M^{3}}\mathcal{H}\pi ^{\prime 3}\right) \psi +\left( \frac{3c_{3}}{%
			a^{2}M^{3}}\pi ^{\prime 3}-6M_{\text{Pl}}^{2}\mathcal{H}\right) \theta ^{\prime } 
		\nonumber \\
		&=&a^{2}\delta \rho _{m}-\frac{c_{3}}{a^{2}M^{3}}\pi ^{\prime 2}\nabla
		^{2}\delta \pi -\left( c_{2}\pi ^{\prime }-\frac{9c_{3}}{a^{2}M^{3}}\mathcal{%
			H}\pi ^{\prime 2}\right) \delta \pi ^{\prime },  \label{00}
	\end{eqnarray}%
	where $\bar{\rho}_{m}=\bar{\rho}_{c}+\bar{\rho}_{b}$, $\text{ \ }$ $\delta \rho _{m} =\bar{\rho}_{m}\delta _{m} 
	=\bar{\rho}_{c}\delta _{c}+\bar{\rho}_{b}\delta _{b}$,$\text{ \ }$  
	$\delta _{c}={\delta \rho _{c}}/{\bar{\rho}_{c}}$ and 
	$\delta _{b}={\delta \rho _{b}}/{\bar{\rho}_{b}}$.
	Substituting Friedmann's equation (\ref{f1}) into (\ref{00}), we obtain:
	\begin{eqnarray}
		&&2M_{\text{Pl}}^{2}\nabla ^{2}\theta -\left( 6M_{\text{Pl}}^{2}\mathcal{H}^{2}+c_{2}\pi
		^{\prime 2}-\frac{12c_{3}}{a^{2}M^{3}}\mathcal{H}\pi ^{\prime 3}\right) \psi
		+\left( -6M_{\text{Pl}}^{2}\mathcal{H+}\frac{3c_{3}}{a^{2}M^{3}}\pi ^{\prime
			3}\right) \theta ^{\prime }  \nonumber \\
		&=&a^{2}\delta \rho _{m}-\frac{c_{3}}{a^{2}M^{3}}\pi ^{\prime 2}\nabla
		^{2}\delta \pi -\left( c_{2}\pi ^{\prime }-\frac{9c_{3}}{a^{2}M^{3}}\mathcal{%
			H}\pi ^{\prime 2}\right) \delta \pi ^{\prime }.
		\label{per eq Einst 00}
	\end{eqnarray}
The $\left( 0,i\right)$  component of Einstein equations is:
	\begin{equation}
		\partial _{i}\left[ 2M_{\text{Pl}}^{2}\theta ^{\prime }+\left( 2M_{\text{Pl}}^{2}\mathcal{H%
		}^{2}-\frac{c_{3}}{a^{2}M^{3}}\pi ^{\prime 3}\right) \psi +\left( c_{2}\pi
		^{\prime }-\frac{3c_{3}}{a^{2}M^{3}}\mathcal{H}\pi ^{\prime 2}\right) \delta
		\pi +\frac{c_{3}}{a^{2}M^{3}}\pi ^{\prime 2}\delta \pi ^{\prime }\right]
		=-a^{2}\bar{\rho}_{m}\upsilon _{i},
	\end{equation}%
	and the $\left( i,j\right)$ components leads to:
	\begin{eqnarray}
		 && M_{\text{Pl}}^{2}  \partial ^{i}\partial _{j}\left(\theta -\psi \right) 
		+\left[ 2M_{\text{Pl}}^{2}\theta ^{\prime \prime } + 4M_{\text{Pl}}^{2}\mathcal{H\theta }^{\prime} \right.
		\nonumber\\
		&+&\left( 2M_{\text{Pl}}^{2}\mathcal{H}^{2}+4M_{\text{Pl}}^{2}\mathcal{H}^{\prime}
		-c_{2}\pi ^{\prime 2}+\frac{4c_{3}}{a^{2}M^{3}}\pi ^{\prime 2}\left( 
		\mathcal{H}\pi ^{\prime }-\pi ^{\prime \prime }\right) \right) \psi +\left( 2M_{\text{Pl}}^{2}\mathcal{H-}\frac{c_{3}}{a^{2}M^{3}}\pi ^{\prime
			3}\right) \psi ^{\prime }
		\nonumber \\
		&+&\left. M_{\text{Pl}}^{2}\nabla ^{2}\left( \psi -\theta \right)
		+\left( c_{2}\pi ^{\prime }+\frac{c_{3}}{a^{2}M^{3}}\pi ^{\prime }\left(
		2\pi ^{\prime \prime }-3\mathcal{H}\pi ^{\prime }\right) \right) \delta \pi
		^{\prime }+\frac{c_{3}}{a^{2}M^{3}}\pi ^{\prime 2}\delta \pi ^{\prime \prime
		}\right] \delta _{j}^{i}=0.
	\end{eqnarray}%
	\newline
As usual we separate this equation into  trace and traceless parts as:
	\begin{eqnarray}
		2M_{\text{Pl}}^{2}\theta ^{\prime \prime }&+&4M_{\text{Pl}}^{2}\mathcal{H}{\theta }^{\prime
		}+\left( 2M_{\text{Pl}}^{2}\mathcal{H}^{2}+4M_{\text{Pl}}^{2}\mathcal{H}^{\prime
		}-c_{2}\pi ^{\prime 2}+\frac{4c_{3}}{a^{2}M^{3}}\pi ^{\prime 2}\left( 
		\mathcal{H}\pi ^{\prime }-\pi ^{\prime \prime }\right) \right) \psi 
		^{\prime }
		\nonumber \\
		&+&\left(
		2M_{\text{Pl}}^{2}\mathcal{H-}\frac{c_{3}}{a^{2}M^{3}}\pi ^{\prime 3}\right) \psi +\frac{2}{3}M_{\text{Pl}}^{2}\nabla ^{2}\left( \psi -\theta \right) \nonumber\\
		&=&-\left( c_{2}\pi ^{\prime }+\frac{c_{3}}{a^{2}M^{3}}\pi ^{\prime }\left(
		2\pi ^{\prime \prime }-3\mathcal{H}\pi ^{\prime }\right) \right) \delta \pi
		^{\prime }-\frac{c_{3}}{a^{2}M^{3}}\pi ^{\prime 2}\delta \pi ^{\prime \prime
		},
	\end{eqnarray}
	\begin{equation}
		\left( \partial ^{i}\partial _{j}-\frac{1}{3}\delta _{j}^{i}\nabla
		^{2}\right) \left( \theta -\psi \right) =0.
	\end{equation}%
The later equation gives $\theta =\psi$, because of the absence of anisotropic stress.
	
	\subsection{Quasi static approximation}
	It is convenient to work in Fourier space, where the scalar modes are expanded as:
	\begin{equation}
		\Phi  \sim\int
		\Phi _{{k}}\exp \left( i\vec{k}.\vec{x}\right) d^{3}k.
	\end{equation}
	Since matter perturbations evolve on spatial scales much smaller than of the Hubble horizon 	($k \gg a H$), we use the so called quasi-static approximation on sub-horizon scales (QSA) where $\left\vert \Phi ^{\prime \prime }\right\vert
	\lesssim \mathcal{H}\left\vert \Phi ^{\prime }\right\vert \ll
	k^{2}\left\vert \Phi \right\vert $. Then Eqs.(\ref{per_delta_dm}), (\ref{per eq ch}) and (\ref{per eq Einst 00}) simplify and become:
	\begin{equation}
		\delta _{c}^{\text{ }\prime }-k^{2}u_{c}-3\theta ^{\prime }=\frac{a}{\bar{%
				\rho}_{c}}(\bar{Q}_{c}\psi +\delta Q_{c}-\bar{Q}_{c}\delta _{c}),\text{ 
		}  \label{delta_Fourier}
	\end{equation}%
	\begin{equation}
		k^{2}\theta =\frac{-a^{2}}{2M_{\text{Pl}}^{2}}\left( \bar{\rho}_{b}\delta _{b}+\bar{%
			\rho}_{c}\delta _{c}\right) -\frac{c_{3}}{2a^{2}M_{\text{Pl}}^{2}M^{3}}\pi
		^{\prime 2}k^{2}\left( \delta \pi \right),  \label{theta}
	\end{equation}%
	\begin{equation}
		\left( c_{2}-\frac{2c_{3}}{a^{2}M^{3}}\left( \pi ^{\prime \prime }+\mathcal{H%
		}\pi ^{\prime }\right) \right) \pi ^{\prime 2}k^{2}\left( \delta \pi \right)
		-\frac{c_{3}}{a^{2}M^{3}}\pi ^{\prime 4}k^{2}\theta =-a^{3}\delta Q_{\left(x\right) \text{ }}^{\text{ }}\pi^{\prime }. \label{delta_pi}
\end{equation}%
Solving Eqs.(\ref{theta}-\ref{delta_pi}) we obtain:
	\begin{equation}
		\delta \pi =-\frac{\left( \bar{\rho}_{b}\delta _{b}+\bar{\rho}_{c}\delta
			_{c}\right) c_{3}\pi ^{\prime 4}+2M_{\text{Pl}}^{2}M^{3}a^{3}\delta Q_{\left(
				{x}\right) \text{ }}^{\text{ }}\pi ^{\prime }}{2M_{\text{Pl}}^{2}M^{3}k^{2}\left[
			\left( c_{2}-\frac{2c_{3}}{a^{2}M^{3}}\left( \pi ^{\prime \prime }+\mathcal{H%
			}\pi ^{\prime }\right) \right) \pi ^{\prime 2}+\frac{c_{3}^{2}}{%
				2a^{4}M_{\text{Pl}}^{2}M^{6}}\pi ^{\prime 6}\right] }.  \label{per_delta_pi}
	\end{equation}%
	Differentiating Eq.(\ref{delta_Fourier}) with respect to the conformal time, and using (\ref{per_dmi}), (\ref{delta_Fourier}) and (\ref{theta}), we  get:
	\begin{eqnarray}
		\delta _{c}^{\prime \prime }+\mathcal{H}\delta _{c}^{\prime } &=&\frac{%
			a^{2}}{2M_{\text{Pl}}^{2}}\left( \bar{\rho}_{b}\delta _{b}+\bar{\rho}_{c}\delta
		_{c}\right) +\frac{c_{3}}{2a^{2}M_{\text{Pl}}^{2}M^{3}}\pi ^{\prime 2}k^{2}\left(
		\delta \pi \right) +\frac{\mathcal{H}a}{\bar{\rho}_{c}}(\delta Q_{c}-\bar{Q%
		}_{c}\delta _{c})+  \nonumber \\
		&&\left[ \frac{a}{\bar{\rho}_{c}}(\delta Q_{c}-\bar{Q}_{c}\delta _{c})%
		\right] ^{\prime }+\frac{a}{\bar{\rho}_{c}}k^{2}\left[ \bar{Q}_{c}\left(
		u-u_{c}\right) +f_{c}\right].  \label{delta_dm''}
	\end{eqnarray}%
	Following  same steps we obtain the evolution of the baryon matter perturbation%
	\begin{equation}
		\delta _{b}^{\prime \prime }+\mathcal{H}\delta _{b}^{\prime }=\frac{a^{2}}{%
			2M_{\text{Pl}}^{2}}\left( \bar{\rho}_{b}\delta _{b}+\bar{\rho}_{c}\delta
		_{c}\right) +c_{3}\frac{\pi ^{\prime 2}k^{2}}{2a^{2}M_{\text{Pl}}^{2}M^{3}}
		\delta \pi.  \label{delta_b"}
	\end{equation}
	 A notable observation from this outcome is the impact  exerted by the coupling between dark energy (DE) and dark matter (DM) on the baryon density perturbation.
	\subsection{Covariant interaction term}
	
	As already discussed above, the simplest
	physical choice for the DE-DM interaction is to assume that there is no momentum transfer in
	the rest frame of neither DM nor DE \cite{valiviita2008large}. Then we choose 
	\begin{equation}
		Q_{c}^{\mu }=Q_{c}u_{c}^{\mu }, \label{choix Q}
	\end{equation}%
		such that Eqs.(\ref{Q0}-\ref{Qi}) read as 
		\begin{equation}
		Q_{c}^{0}=a^{-1}\left( \bar{Q}_{c}\left( 1-\psi \right) +\delta
		Q_{c}\right),
	\end{equation}
		\begin{equation}
		Q_{c}^{i}=a^{-1}\partial ^{i}\left( \bar{Q}_{c}u_{\text{ \ }c}^{\left(
			1\right) }\right),
	\end{equation}%
	and
	\begin{equation}
		f_{c}=Q_{c}\left( u_{c}-u\right) =-f_{x}.
	\end{equation}%
		The preceding treatment of perturbations is completely general. Let us specify 
	the form of the energy-momentum transfer rate as \cite{valiviita2008large}:
		\begin{equation}
		Q_{c}=-Q_{x}=-\alpha \frac{\mathcal{H}}{a}\rho _{x}.
		\label{delta_Q_dm}
	\end{equation}%
	Substituting Eqs. (\ref{delta_Q_dm}) and (\ref{delta_rho_de}) in (\ref%
	{per_delta_pi}) we obtain
	\begin{equation}
		\delta \pi =-\frac{\left( \bar{\rho}_{b}\delta _{b}+\bar{\rho}_{c}\delta
			_{c}\right) c_{3}\pi ^{\prime 4}}{2M_{\text{Pl}}^{2}M^{3}k^{2}\left[ \left( c_{2}-%
			\frac{2c_{3}}{a^{2}M^{3}}\left( \pi ^{\prime \prime }+\mathcal{H}\pi
			^{\prime }\right) \right) \pi ^{\prime 2}+ \frac{\alpha c_{3}}{M^{3}}\mathcal{H}%
			\pi ^{\prime 3}+\frac{c_{3}^{2}}{%
				2a^{4}M_{\text{Pl}}^{2}M^{6}}\pi ^{\prime 6}\right] }.  \label{delta pi}
	\end{equation}%
	Inserting (\ref{delta pi}) in (\ref{delta_dm''}) we obtain
	\begin{equation}
		\delta _{c}^{\prime \prime }+\left( 1-\alpha \gamma \right) \mathcal{H}%
		\delta _{c}^{\prime } =4\pi Ga^{2}\left(\bar{\rho}_{b}\delta _{b}+\bar{\rho}_{c}\delta _{c}\right)\left( 1-\frac{%
			c_{\text{ }3}^{2}\pi ^{\prime 4}}{2c_{2}a^{4}M_{\text{Pl}}^{2}M^{6}\lambda }\right)
		+  { \xi}\bar{\rho}_{c}\delta _{c},%
		\label{per delta dm''}
	\end{equation}%
	\begin{equation}
		\delta _{b}^{\prime \prime }+\mathcal{H}\delta _{b}^{\prime }=4\pi
		Ga^{2}\left( \bar{\rho}_{b}\delta _{b}+\bar{\rho}_{c}\delta _{c}\right) \left(1-%
		\frac{c_{\text{ }3}^{2}\pi ^{\prime 4}}{2c_{2}a^{4}M_{\text{Pl}}^{2}M^{6}\lambda }\right),
		\label{per delta'' b}
	\end{equation}%
	where 
	\begin{equation}
		\lambda =1-\frac{2c_{3}}{c_{2}a^{2}M^{3}}\left( \pi ^{\prime \prime }+%
		\mathcal{H}\pi ^{\prime }\right) +\frac{c_{3}^{2}}{2c_{2}a^{4}M_{\text{Pl}}^{2}M^{6}%
		}\pi ^{\prime 4}+ \frac{\alpha c_{3}}{c_{2}M^{3}}\mathcal{H}\pi ^{\prime },
	\end{equation}%
	and%
	\begin{equation}
		\xi =\frac{\alpha}{\bar{\rho}_{c}}\left( \mathcal{H}^{2}\gamma +\mathcal{H}%
		^{\prime }\gamma +\mathcal{H}{\gamma }^{\prime }\right),\,\,%
		\gamma =\text{\ }\frac{\bar{\rho}_{x}}{\bar{\rho}_{c}}.
	\end{equation}%
	Clearly, the growth rate of matter in  $\Lambda \text{CDM}$ is recovered for  $c_{3}=0$ and $\alpha =0$.
	
	For numerical purpose it is useful to work in terms of the time $N=\ln (a)$ and the dimensionless  variables of Sec.(\ref{dyn.sys}). Adopting the notation $\frac{df}{dN}$ $=f'$, straightforward calculations lead to the relations:
		\begin{equation}
		\delta _{c}^{''}+\left( 2-\alpha \gamma +\epsilon_H\right) \delta _{c}^{'} =\frac{3}{2}\left(G_{cc}\bar{\Omega}_{cc}\delta _{c}+G_{cb}\bar{\Omega}_{b}\delta _{b}\right),
		\label{per_delta_dm,,}
	\end{equation}%
	\begin{equation}
		\delta _{b}^{''}+\left( 2+\epsilon_H\right)
		\delta _{b}^{'}=\frac{3}{2}\left(G_{bb} \bar{\Omega}_{b}\delta _{b}+G_{bc}\bar{\Omega}%
		_{c}\delta _{c}\right),  \label{per_delta_b,,}
\end{equation}%
where the effective gravitational constants are given by:
	\begin{equation}
		G_{bb}=1-\left( \frac{\alpha -6}{\alpha -3}\right) ^{2}\frac{r_{1}r_{2}}{12\lambda }%
		, \,\, G_{cc}=G_{bb}+ \xi,\,\,G_{cb}=G_{bc}=G_{bb},
	\end{equation} 
	\begin{eqnarray}
			\lambda &=&1-\frac{\left( \alpha -6\right) }{3\left( \alpha -3\right) r_{1}}%
		\left( \epsilon_{\pi}+2\right) +\left( \frac{\alpha -6}{%
			\alpha -3}\right) ^{2}\frac{r_{1}r_{2}}{12}+\frac{\alpha \left( \alpha
			-6\right) }{\left( \alpha -3\right) }\frac{1}{6 r_{1}},\\
		\xi &=&\frac{2}{3}\frac{\alpha
			\gamma }{\bar{\Omega}_{c}}\left( 2+\epsilon_H+\alpha \left( 1+\gamma
		\right) -3\omega _{x}\right),
	\end{eqnarray}%
and $\epsilon_H$, $\epsilon_{\pi}$ are given by Eqs.\eqref{epsilonH} and \eqref{epsilonpi}.

		As a result of unequal couplings for DM and baryons, we expect the existence of a bias between baryons and DM. We
	study this in the DM dominated scenario,$\vert\Omega_{c}\delta_{c}\vert \gg\vert\Omega_{b}\delta_{b}\vert $, and define a constant
	bias $b$, by $\delta_b = b\, \delta_c$. We can easily determine the bias by writing Eqs. \eqref{per_delta_dm,,} and \eqref{per_delta_b,,} in terms of
	the DM growth parameter $f_c = d \ln\left(\delta_c\right)/dN$. Indeed, we easily find 
	\begin{equation}
		b=\frac{2 f_{c}\left(2-\alpha\gamma+\epsilon_H+\delta_c\right)+3 \left(G_{bc}-G_{cc}\right)\Omega_c}%
		{2 f_{c}\left(2+\epsilon_H+\delta_c\right)}.
	\end{equation}
	In the limit $\alpha\rightarrow 0$, we have $G_{cc}\rightarrow  G_{bc}$,  and then $b=1$.

	\section{Methodology and data\label{sec:Methodology and data}}
	
	In this section, we establish observational constraints on the cubic covariant Galileon field coupled to dark matter through the application of a Markov Chain Monte Carlo (MCMC) integration using the Metropolis-Hastings algorithm. We use the recent compilation of redshift space distortion (RSD) datasets, with a specific emphasis on the lowermost 21 data points \cite{perivolaropoulos,perivolaropoulos2} and the model-independent observational Hubble  dataset (OHD) \cite{Jimenez_2002}  listed  in the Table in appendix \ref{appendixA}. We also utilize combined data from Cosmic Microwave Background (CMB) observations and Baryon Acoustic Oscillations (BAO) based on the Planck 2015 distance prior and a set of BAO measurements \cite{Santos_2016}. Additionally, we include the efficient  catalog of six measurements of $E^{-1}(z)=\left(H(z)/H_{0}\right)^{-1}$, whose  efficacy was validated for spatially flat cosmologies \cite{Riess_2018}. These measurements were obtained from a compressed catalog comprising 1052 Type Ia supernovae (SN), supplemented by the MCT set containing 15 SN at redshifts $z>1$. The resulting combined likelihood function is formulated as:
\begin{equation}
\mathcal{L}({\cal{P}})=\mathcal{L}_{\text{BAO/CMB}} \times\mathcal{L}_{\text{RSD}} \times \mathcal{L}_{\text{OHD}} \times  \mathcal{L}_{\text{SN}},
\end{equation}
		where ${\cal{P}}$ is the vector of model parameters over which the 	MCMC integration is performed. 
		
We perform our analysis for the full exact model labeled ICG  	and the model obtained through the tracker solution labeled TICG,  obtained from the ICG by setting $r_{1}=1$ in the dynamical equations. For the ICG model the parameter vector  is given by ${\cal{P}}=\left(\alpha,\,r_{1i},\,r_{2i},\,h,\,\sigma_{8}\right)$, 	while in the TICG model it is  ${\cal{P}}=\left(\alpha,\,r_{2i},\,h,\,\sigma_{8}\right)$, where $r_{1i}$ and $r_{2i}$ are the starting values deep in the radiation epoch. 	The ICG model has one more parameter than the TICG model and  is expected to be more penalized by Bayesian selection. All these free parameters are explored within the range of the conservative flat priors listed in
Table I. Without using any fiducial cosmology to correct the $f\sigma_{8}$ measurements, we confront our findings at $68\%$, $95\%$ confidence limits (CL) with the $\Lambda\textrm{CDM}$ model. All the constraints presented below are derived using \texttt{getdist}\footnote{\url{https://github.com/cmbant/getdist}}~\cite{Lewis:2019xzd}. 
		 \begin{table}[H]
	   \begin{centering}
	     \begin{tabular}{c c c c}
	 
	 \hline \hline 
	 Parameter & ICG  & TICG & $\Lambda$CDM \\ 
	 \hline
	 $\alpha$ & $\left[-1,0\right]$ & $\left[-1,0\right]$&$-$ \\ 
	 
	 $r_{1}^{(0)}$ & $\left[0,1\right]\times 10^{-12}$ & $-$ &$-$\\ 
	  
	 $r_{2}^{(0)}$ & $\left[0,3\right]\times10^{-17}$ & $\left[0,2\right]\times10^{-60}$ &$-$ \\ 
	 $\Omega_{m}^{(0)}$ & $-$ & $-$ &$ \left[0.1,0.4\right]$\\
	 $h$ & $\left[0.6,0.9\right]$ & $\left[0.6,0.9\right]$& $\left[0.6,0.9\right]$\\ 
	 
	 $\sigma_8$ & $\left[0.5, 1.5\right]$ & $\left[0.5, 1.5\right]$ &$\left[0.5, 1.5\right]$\\ 
	 \hline \hline
	   \end{tabular} 
	   \caption{Flat priors on the cosmological parameters.}\label{priors}
	  \end{centering}
	 \end{table}
In addition, we evaluate the goodness of fit using  the corrected frequentist Akaike Information Criterion (AIC) \cite{1974ITAC} defined by:
\begin{equation}
\textrm{AIC}_c=\textrm{AIC}+\frac{2 k\left(k+1\right)}{n-k+1},
\end{equation}
where $\textrm{AIC}$ is the standard  criterion, given by:
\begin{equation}
\textrm{AIC}=-2 \ln \mathcal{L}_{\textrm{max}}+2 k,
\end{equation} 
where $k$ is the number of estimated parameters in the model, and $n$  the number of data points. The $\textrm{AIC}$ penalizes model complexity by adding twice the number of free parameters to the $\chi^2$ value, thereby favoring simpler models when their $\chi^2$ values are comparable. The corrected $\textrm{AIC}$ ($\textrm{AIC}_c$) further refines this approach by including a correction term that accounts for small sample sizes, helping to prevent overfitting in cases where the number of data points is limited. 
The best-fit model is determined by minimizing the $\textrm{AIC}_c$ score: lower $\textrm{AIC}_c$ values indicate a better fit. Deviations between models are assessed using Jeffreys' scale, where a difference $\Delta\textrm{AIC}_c>5\left(10\right)$ is considered strong (decisive) evidence against the model with the higher $\textrm{AIC}_c$ score. 

\section{Results and discussion\label{sec:results and discussion}}
\begin{figure}[H]
		\begin{center}
			\includegraphics[width=9cm,height=8cm]{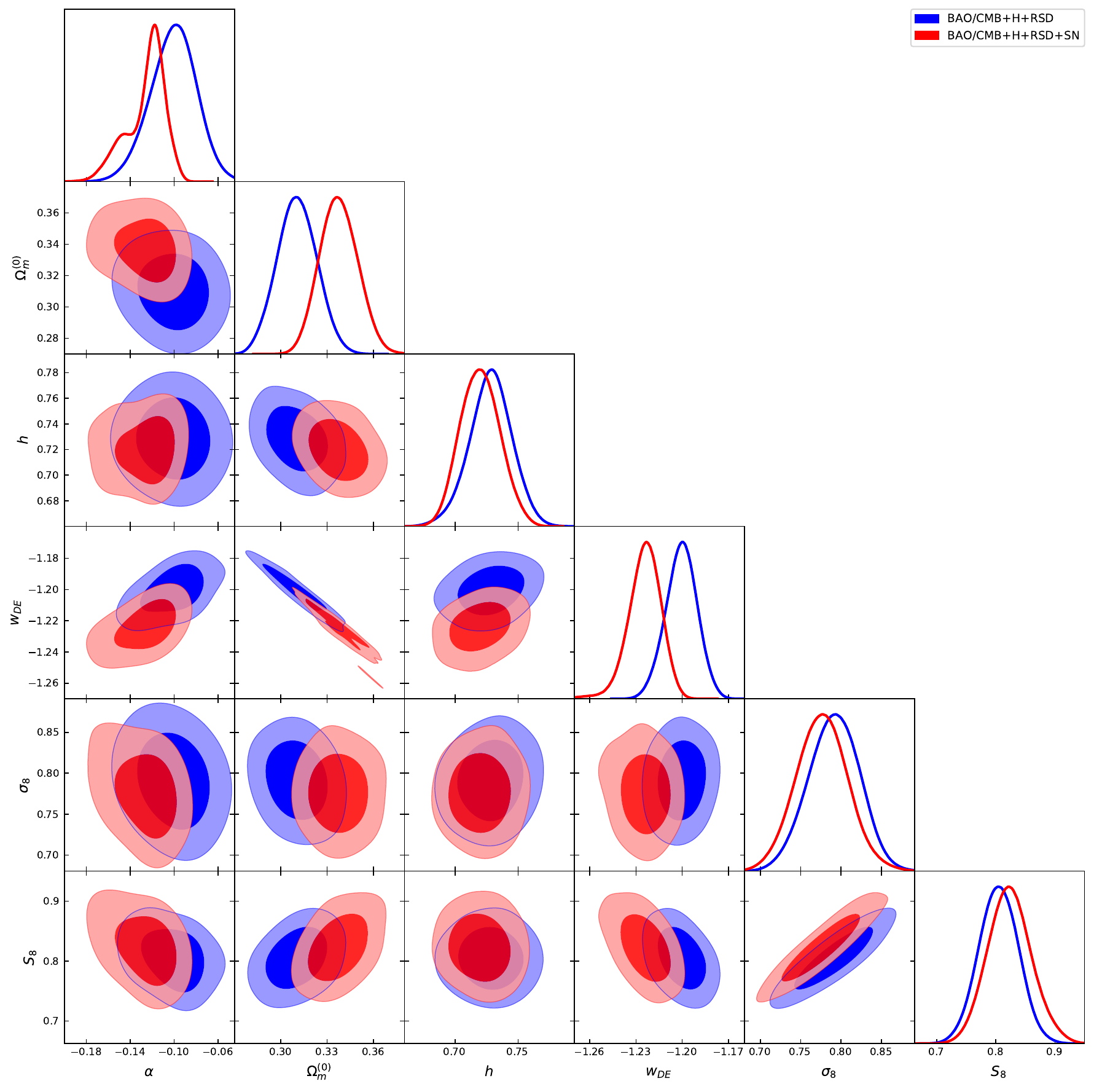}
			\caption{\label{contoursHGBAOCMB} One-dimensional marginalized posterior distributions and two-dimensional joint contours at $1 \sigma$ and $2 \sigma$ confidence levels for the BAO/CMB+H+RSD (light blue) and BAO/CMB+H+RSD+SN (orange) datasets for the ICG  model.}
		\end{center}
	\end{figure}

\begin{figure}[H]
	\centering
	\includegraphics[scale=0.34]{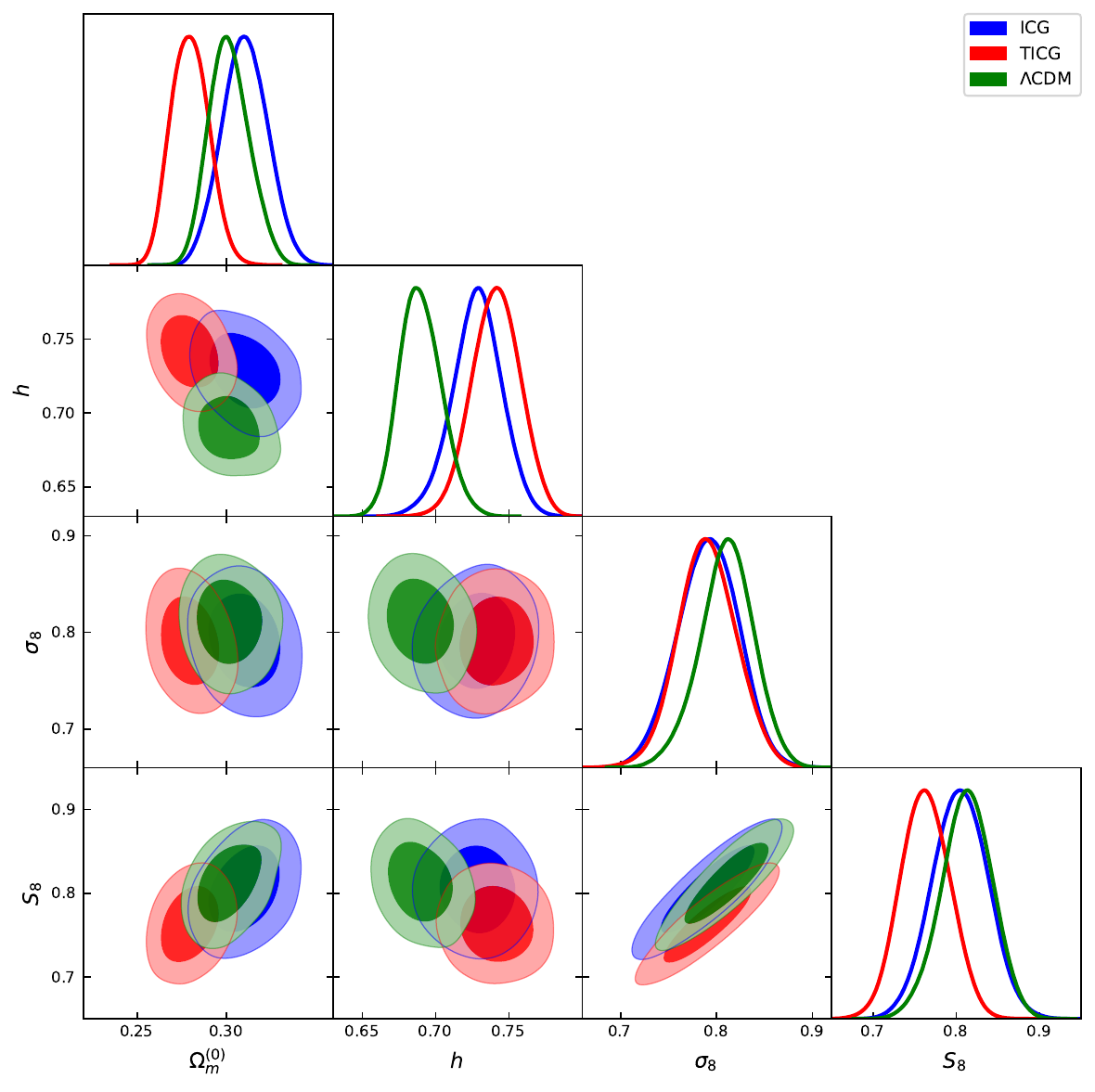}
	\includegraphics[scale=0.34]{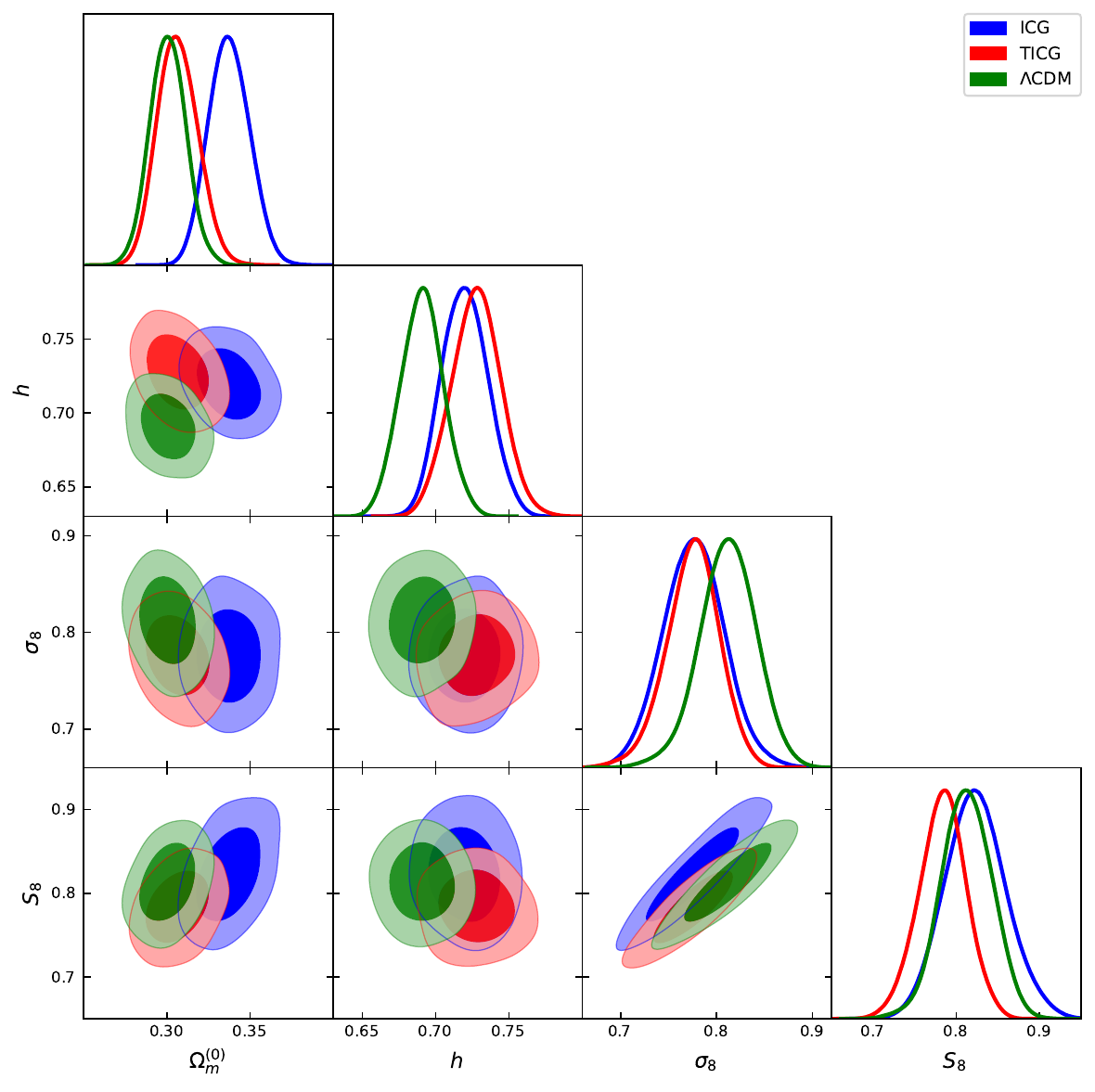}
	\caption{1-D marginalized posterior and 2-D joint contours at $1 \sigma$ and $2 \sigma$ CL  of various models for dataset  BAO/CMB+H+RSD (left panel), and the entire dataset BAO/CMB+H+ RSD+SN (right panel).}
	\label{contoursall}
\end{figure}

The cosmological analysis involves fitting the ICG and TICG models to two dataset combinations: BAO/CMB+H+RSD, and this combination augmented with the SN dataset. For comparison, we also include the analysis of the $\Lambda$CDM model.  The resulting parameter constraints at the $68\%$ CL, the $\chi^2$ statistics and $\textrm{AIC}_c$ are summarized in Tables \ref{t2} and \ref{t3}. In Figures \ref{contoursHGBAOCMB} $\sim$ \ref{contoursall} we show the 1-D marginalized posterior distributions and 2-D joint contours at $1 \sigma$ and $2 \sigma$ CL.

The results outlined in Table \ref{t2} for BAO/CMB+H+RSD datasets offer significant insights. The initial conditions $r_{i1}$ and $r_{i2}$ reported at $68\%$ CL are obtained for 
$\Omega_r^{(i)}=0.9999929$ at $z=4.85 \times 10^8$. 
 We observe that the mean value  at  $68\%$ CL of $H_0$ reported for the ICG and TICG, respectively, are higher than that reported for $\Lambda$CDM. This is significant in light of addressing the tension between the Planck-inferred value $H_0=67.4\pm 0.5$ km/s/Mpc \cite{aghanim2020planck} and the locally measured value $H_0=74.03\pm 1.42$ km/s/Mpc by Riess et al. \cite{Riess_2022,2023JCAP}, indicating a reduction in tension to less than $1\sigma$. Our results for $\sigma_8$ indicate that the mean values for the ICG and TICG are about $0.5\sigma$ lower than that of  $\Lambda$CDM model. It can be said that the  $\Lambda$CDM and the ICG and TICG models are in perfect agreement with the results of low-redshift observations and the $\sigma_8$ value inferred by Planck. However, the ICG model produced an $S_8$ value very close to that reported by the concordance model, thereby maintaining the tension with LSS observations by DES-Y3 \cite{2022PhRvD.105b3520A}. In contrast, the TICG model yielded a much lower value that aligns well with LSS observations, at $0.4\sigma$ from DES-Y3 \cite{2022PhRvD.105b3520A}. The discrepancy between the two solutions arises primarily because  ICG  allows for a higher value of $\Omega_m^{(0)}$, while  TICG  accommodates a smaller value compared to that of  $\Lambda$CDM. The findings for the TICG model are consistent with the observation that the quantity $\Omega_m^{(0)}h^2$ is tightly constrained by Planck. Additionally, the existence of energy transfer from DE to DM  suggests that the contribution from DM density needs to be reduced. Specifically, for negative $\alpha$, the DM density receives an additional contribution proportional to the DE density.
 
\begin{figure}[H]
		\begin{center}
			\includegraphics[width=9cm,height=8cm]{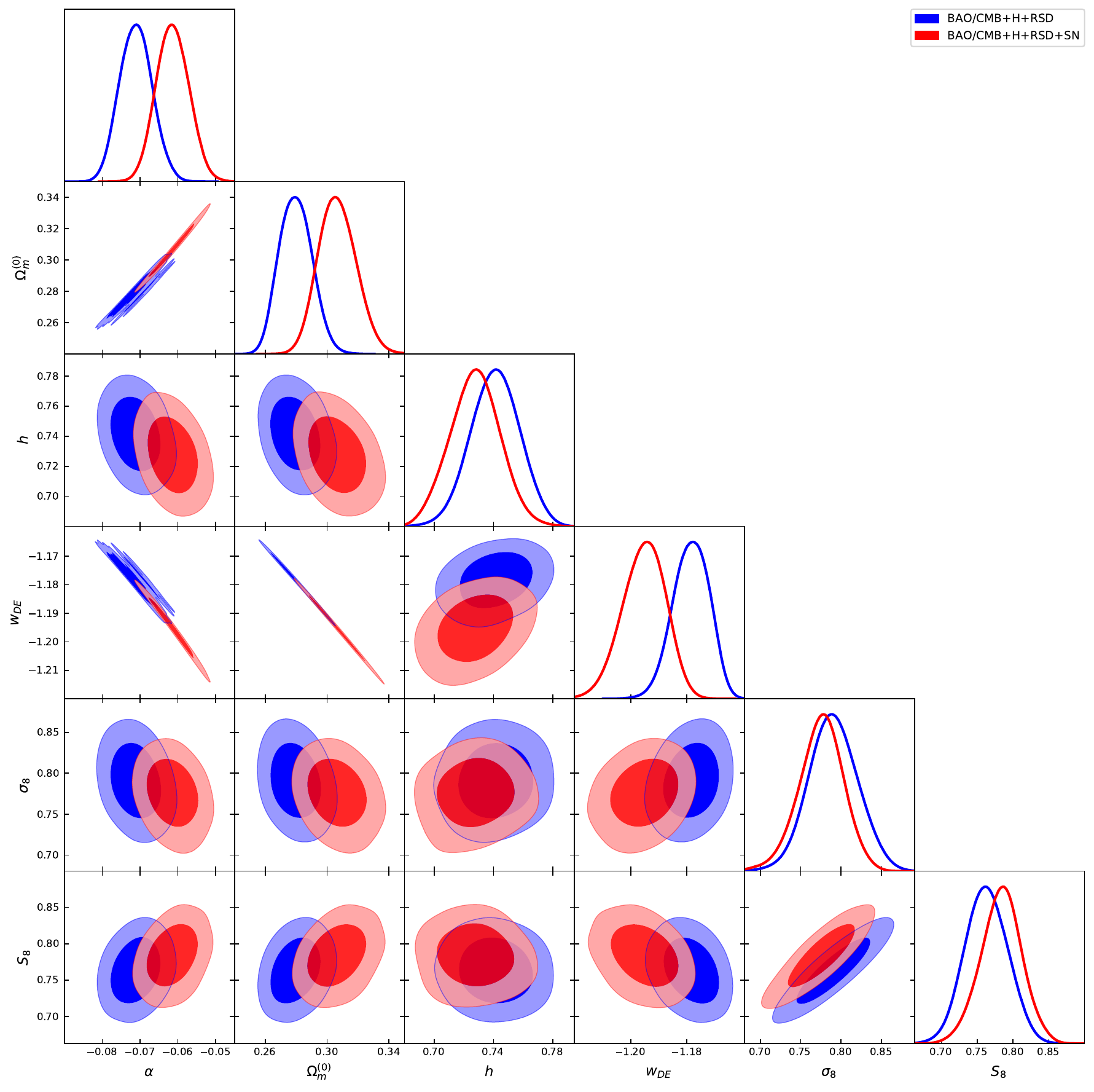}
			\caption{\label{contoursHGBAOCMBSN}One-dimensional marginalized posterior distributions and two-dimensional joint contours at $1 \sigma$ and $2 \sigma$ confidence levels for the BAO/CMB+H+RSD (light blue) and BAO/CMB+H+RSD+SN (orange) datasets for the TICG  model.}
		\end{center}
	\end{figure}

Furthermore, the present day DE EoS  lies within the phantom regime at $68\%$ CL, with $w_0=−1.2 \pm 0.01$ for ICG  and $w_0=−1.18^{+0.0069}
_{−0.0060}$ for  TICG, respectively. These values correlate with  higher expansion rates, as illustrated in Fig. \ref{contoursHGBAOCMB}. This is true if we impose consistency with CMB observations. Indeed, the positions of the acoustic peaks in the CMB are influenced by the overall distance that light can travel from the surface of last scattering to us. When $w$ is allowed to vary into the phantom region, the faster expansion due to $w<-1$  requires an increased $H_0$ to keep the acoustic peak positions unchanged. Moreover, the DM-DE coupling parameter exhibits mean values given by $\alpha=-0.101_{-0.019}^{+0.022}$ (ICG) and $\alpha=-0.071\pm 0.0043$ (TICG) at $68\%$ CL. This complex interplay between $\alpha,\,w_0$ and $H_0$ can be observed in figures \ref{contoursHGBAOCMB} and \ref{contoursHGBAOCMBSN},  where a strong degeneracy between $\alpha$ and $w_{DE}$ dominate the one between $\alpha$ and $H_0$ in the ICG model.

Similarly, according to the results in Table \ref{t3}, the parameter space becomes more constrained with the inclusion of SN data. For the entire dataset, including BAO/CMB+H+RSD and SN, the best fit initial conditions allowed for $\Omega_r^{(i)}=0.9999929$ at $z=4.85\times 10^8$ are slightly reduced for ICG and TICG compared to the experiment with the BAO/CMB+H+RSD dataset. Specifically, $H_0=72\pm 1.5$ km/s/Mpc and $H_0=72.8\pm 1.7$ km/s/Mpc at a $68\%$ CL for ICG and TICG, respectively. These values remain larger than those predicted by the $\Lambda$CDM model, reducing the tension with the value reported by SH0ES (R19) \cite{Riess_2022,2023JCAP} to $0.5\sigma$ and $0.1\sigma$ for ICG and TICG, respectively.

Additionally, the inclusion of SN data leads to a decrease in $\sigma_8$ for the ICG and TICG models, while it causes an increase in $\sigma_8$ for the $\Lambda$CDM model. Concurrently, an increase in $\Omega_{m}^{(0)}$ is observed for both ICG and TICG, similar to the scenario observed in the non-interacting covariant Galileon model \cite{2010PhRvD82l4054N}. As a result, the tension in $S_8$ for the ICG model increases to approximately $1\sigma$, while it is alleviated in the TICG model to around $0.2\sigma$ from DES-Y3 data.

Moreover, a non-zero coupling constant, $\alpha=-0.128_{-0.0087}^{+0.0250}$ and $\alpha=-0.0614\pm 0.0043$ at a $68\%$ CL for ICG and TICG, respectively, is also obtained. In the ICG model, the negative increase in $\alpha$ is accompanied by a substantial decrease in $w_0$ and an increase in $\Omega_{0}^{(0)}$, driven by the energy transfer from DE to DM, and a shift in $H_0$ towards smaller values. In contrast, the TICG model shows an increase in the DM-DE coupling, accompanied by an increase in 
$\Omega_{0}^{(0)}$, and a decrease in $w_0$ and $H_0$. Additionally, the significant decrease in the Hubble constant is crucial for maintaining $\Omega_m^{(0)}h^2$ nearly constant. 
However, the inclusion of the SN data does not mitigate the tension with LSS observations; it remains at the same level as observed in the experiment without SN data.

For the two experiments, we observe that ICG exhibits a robust correlation between $\alpha$ and $w_0$, meaning that the phantom regime is enhanced by the DM-DE interaction. In ICG, the resolution of the $H_0$ tension is driven by the phantom regime, while the $S_8$ tension is not resolved and remains at the level observed in $\Lambda$CDM, as seen in Fig. \ref{contoursHGBAOCMB}. For TICG, we observe in Fig. \ref{contoursHGBAOCMBSN} that $\alpha$ and $h$ are anti-correlated, while $\alpha$ and $S_8$ are correlated, explaining why both the $H_0$ and $S_8$ tensions are alleviated within the TICG solution. Finally, we observe that the 1-D posterior distributions feature narrow peaks comfortably situated within the permissible range defined by the priors, suggesting that the outcomes are not overly reliant on the selected parameters.

Moreover, the ICG model is free from the strong degeneracy between the parameters $\alpha$ and $\Omega_m^{(0)}$, observed in  TICG. Importantly, the ICG and TICG models are free from non-adiabatic instabilities at large scales. This is evidenced by the doom factor, expressed as $d=−Q/\left(3\rho_x(1+w_x)\right)$, being consistently negative \cite{Gavela:2009}.

A summary of our results can be drawn from Fig.  \ref{contoursall}. We clearly observe a shift in the current expansion rate towards higher values for both the ICG and TICG models compared to $\Lambda$CDM, alleviating the current tension between early and late-time observations. Specifically, we observe an excess in the current DM density for the ICG model, and a shift of $S_8$ to smaller values for the TICG model, which further alleviates the tension on $S_8$. According to the values of $\textrm{AIC}_c$ in Tables \ref{t2} and \ref{t3}, the ICG model is slightly disfavored compared to $\Lambda$CDM, whereas the TICG model and $\Lambda$CDM are consistent with each other. 
Overall, the inclusion of SN data reveals significant dynamics in the interplay between cosmological parameters, underscoring the robustness of the ICG and TICG models in addressing various cosmological tensions while remaining stable at large scales.

	\begin{table}[H]
		\begin{centering}
		\resizebox{12cm}{!}{
			\begin{tabular*}{14cm}{@{\extracolsep{\fill}}c c c c}
				\hline \hline
				\multicolumn{4}{c}{BAO/CMB+H+RSD} \\
				\hline
				\textbf{\scriptsize{\
						Parameter}} & {\scriptsize{}
$\begin{array}{c}
						\underline{\text{ICG}}\\
						\chi_{min}^{2}=23.0529
					\end{array}$} & {\scriptsize{}$\begin{array}{c}
						\underline{\text{TICG}}\\
						\chi_{min}^{2}=23.1082
					\end{array}$} & {\scriptsize{}$\begin{array}{c}
						\underline{\text{\ensuremath{\Lambda}CDM}}\\
						\chi_{min}^{2}=22.8345
					\end{array}$}
				\tabularnewline
				\hline
				{\scriptsize{}$\alpha$} & {\scriptsize{}$-0.101 (\textcolor{red}{-0.076})_{-0.019}^{+0.022}$} & {\scriptsize{}$-0.0711 (\textcolor{red}{-0.07})\pm 0.0043$} & {\scriptsize{}$-$} 
				\tabularnewline
				{\scriptsize{}$10^{12} r_{1i}$} & {\scriptsize{}$0.093 (\textcolor{red}{0.095})_{-0.065}^{+0.089}$} & $-$ & {\scriptsize{}$-$}
				\tabularnewline
				{\scriptsize{}$10^{18} r_{2i}$} & {\scriptsize{}$7.6 (\textcolor{red}{10.11})_{-1.8}^{+2.8} $} & {\scriptsize{}$\left(16.03 (\textcolor{red}{16.34})_{-0.49}^{+0.41}\right)\times 10^{-43 }$} & {\scriptsize{}$-$} 
				\tabularnewline
				{\scriptsize{}${\Omega_{m}^{(0)}}$} & {\scriptsize{}$\mathbf{0.31 (\textcolor{red}{0.30})\pm 0.013}$} & {\scriptsize{}$\mathbf{0.28 (\textcolor{red}{0.277})_{-0.011}^{+0.010}}$} & {\scriptsize{}$0.3016 (\textcolor{red}{0.302})\pm 0.0112$}
				\tabularnewline
				{\scriptsize{}$h$} & {{\scriptsize{}$0.728 (\textcolor{red}{0.738})\pm 0.017$}} & {\scriptsize{}$0.741 (\textcolor{red}{0.741})\pm 0.016$} & {\scriptsize{}$0.6896 (\textcolor{red}{0.688})\pm 0.0142$}
				\tabularnewline
				{\scriptsize{}$w$} & {{\scriptsize{}$\mathbf{-1.2 (\textcolor{red}{-1.19})\pm 0.01}$}} & {\scriptsize{}$\mathbf{-1.18 (\textcolor{red}{-1.176})_{-0.0060}^{+0.0069}}$} & {\scriptsize{}$-1$}	
				\tabularnewline
				{\scriptsize{}$\sigma_{8}$} & {\scriptsize{}$0.791 (\textcolor{red}{0.784})\pm 0.031$} & {\scriptsize{}$0.79 (\textcolor{red}{0.797})\pm 0.03$} & {\scriptsize{}$0.811 (\textcolor{red}{0.81})\pm 0.0278$} 			
				\tabularnewline
				{\scriptsize{}$S_{8}$} & {\scriptsize{}$\mathbf{0.805 (\textcolor{red}{0.788})\pm 0.033}$} & {\scriptsize{}$\mathbf{0.763 (\textcolor{red}{0.766})\pm 0.03}$} & {\scriptsize{}$\mathbf{0.813 (\textcolor{red}{0.812})\pm 0.043}$}
				\tabularnewline
				\hline
				{\scriptsize{}$\Delta{\chi^2}$} & {\scriptsize{}$0.2184$} & {\scriptsize{}$0.2737$} & {\scriptsize{}$-$}
				\tabularnewline
				{\scriptsize{}$\Delta{\textrm{AIC}_c}$} & {\scriptsize{}$4.8313$} & {\scriptsize{}$2.5423$} & {\scriptsize{}$-$}
				\tabularnewline
				
				\hline\hline
\end{tabular*}}
			\caption{\label{t2}Mean values at 1$\sigma$ CL for the ICG, TICG  and the  $\Lambda$CDM
				models for the combination BAO/CMB+H+RSD. The derived parameters are shown in bold, and in red we show the best fit values.}
		\end{centering}
	\end{table}

	\begin{table}[H]
		\begin{centering}
		\resizebox{12cm}{!}{
			\begin{tabular*}{14cm}{@{\extracolsep{\fill}}c|c|c|c}
				\hline\hline
				\multicolumn{4}{c}{BAO/CMB+H+RSD+SN} \\
				\hline
				\textbf{\scriptsize{Parameter}} & {\scriptsize{}$\begin{array}{c}
						\underline{\text{ICG}}\\
						\chi_{min}^{2}=29.1583
					\end{array}$} & {\scriptsize{}$\begin{array}{c}
						\underline{\text{TICG}}\\
						\chi_{min}^{2}=29.223
					\end{array}$} & {\scriptsize{}$\begin{array}{c}
						\underline{\text{\ensuremath{\Lambda}CDM}}\\
						\chi_{min}^{2}=28.4213
					\end{array}$}
				\tabularnewline
				\hline
				{\scriptsize{}$\alpha$} & {\scriptsize{}$-0.128 (\textcolor{red}{-0.111})_{-0.0087}^{+0.025}$} & {\scriptsize{}$-0.0614 (\textcolor{red}{-0.06})\pm 0.0043 $} & {\scriptsize{}$-$} 
				\tabularnewline
				{\scriptsize{}$10^{12} r_{1i}$} & {\scriptsize{}$0.05 (\textcolor{red}{0.063})_{-0.021}^{+0.019}$} & {\scriptsize{}$-$} & {\scriptsize{}$-$}
				\tabularnewline
				{\scriptsize{}$10^{18} r_{2i}$} & {\scriptsize{}$4.3 (\textcolor{red}{3.9})_{-1.2}^{+2.2}$} & {\scriptsize{}$\left(15.53 (\textcolor{red}{15.73})\pm 0.14\right)\times 10^{-43}$} & {\scriptsize{}$-$} 
				\tabularnewline
				{\scriptsize{}$\Omega_{m}^{(0)}$} & {\scriptsize{}$\mathbf{0.337 (\textcolor{red}{0.331})\pm 0.012}$} & {\scriptsize{}$\mathbf{0.307 (\textcolor{red}{0.308})_{-0.013}^{+0.014}}$} & {\scriptsize{}$0.301 (\textcolor{red}{0.302})\pm 0.0111 $}
				\tabularnewline
				{\scriptsize{}$h$} & {\scriptsize{}$0.72 (\textcolor{red}{0.722}) \pm 0.015 $} & {\scriptsize{}$0.728 (\textcolor{red}{0.728}) \pm 0.017 $} & {\scriptsize{}$0.691 (\textcolor{red}{0.688}) \pm 0.014$}
				\tabularnewline
				{\scriptsize{}$w$} & {\scriptsize{}$\mathbf{-1.224 (\textcolor{red}{-1.22})_{-0.0092}^{+0.0110}}$} & {\scriptsize{}$\mathbf{-1.1953  (\textcolor{red}{-1.196})_{-0.0071}^{+0.0083}}$} & {\scriptsize{}$-1$}
				\tabularnewline
				{\scriptsize{}$\sigma_{8}$} & {\scriptsize{}$0.776 (\textcolor{red}{0.776}) \pm 0.032 $} & {\scriptsize{}$0.775 (\textcolor{red}{0.778})_{-0.025}^{+0.029} $} & {\scriptsize{}$0.812 (\textcolor{red}{0.809}) \pm 029 $} 
				\tabularnewline
				{\scriptsize{}$S_{8}$} & {\scriptsize{}$\mathbf{0.823 (\textcolor{red}{0.816}) \pm 0.037} $} & {\scriptsize{}$\mathbf{0.783 (\textcolor{red}{0.788}) \pm 0.029}$} & {\scriptsize{}$\mathbf{0.812 (\textcolor{red}{0.813})\pm 0.03}$}
				\tabularnewline
				\hline
				{\scriptsize{}$\Delta{\chi^2}$} & {\scriptsize{}$0.737$} & {\scriptsize{}$0.8017$} & {\scriptsize{}$-$}
				\tabularnewline
				{\scriptsize{}$\Delta{\textrm{AIC}_c}$} & {\scriptsize{}$5.35$} & {\scriptsize{}$3.07$} & {\scriptsize{}$-$}
				\tabularnewline
				\hline\hline
\end{tabular*}}
			\caption{\label{t3} Mean values at 1$\sigma$ CL for the ICG, TICG  and the  $\Lambda$CDM
				models for the full dataset BAO/CMB+H+RSD+SN. The derived parameters are shown in bold and the best fit values in red.}
		\end{centering}
	\end{table}

%%%%%%%%%%%%%%%%%%%%%%%%%%%%%%%%%%%%%%%%%%%%%%%%%%%
\begin{figure*}
	\centering
	\includegraphics[scale=0.25]{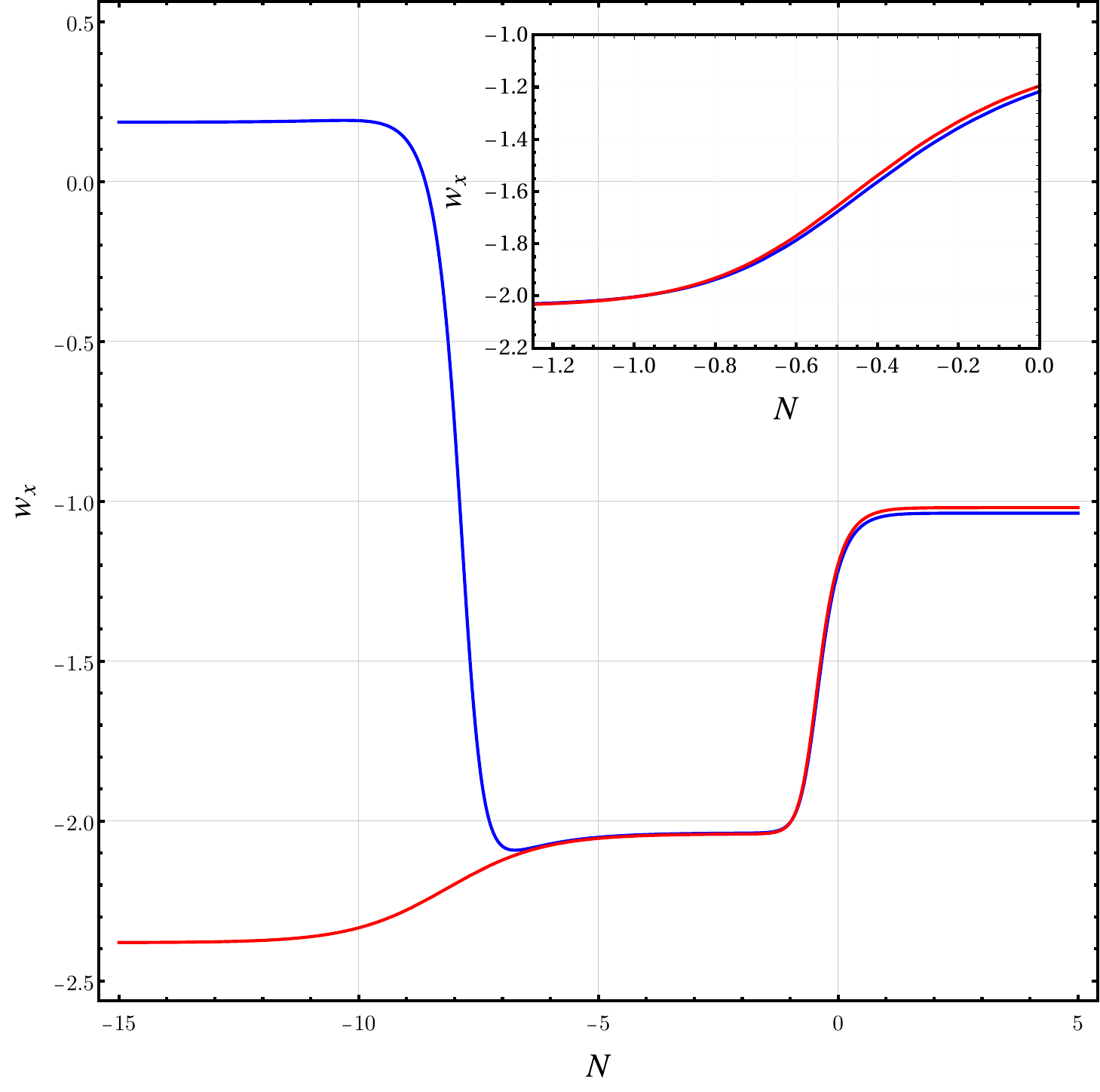}
	\includegraphics[scale=0.25]{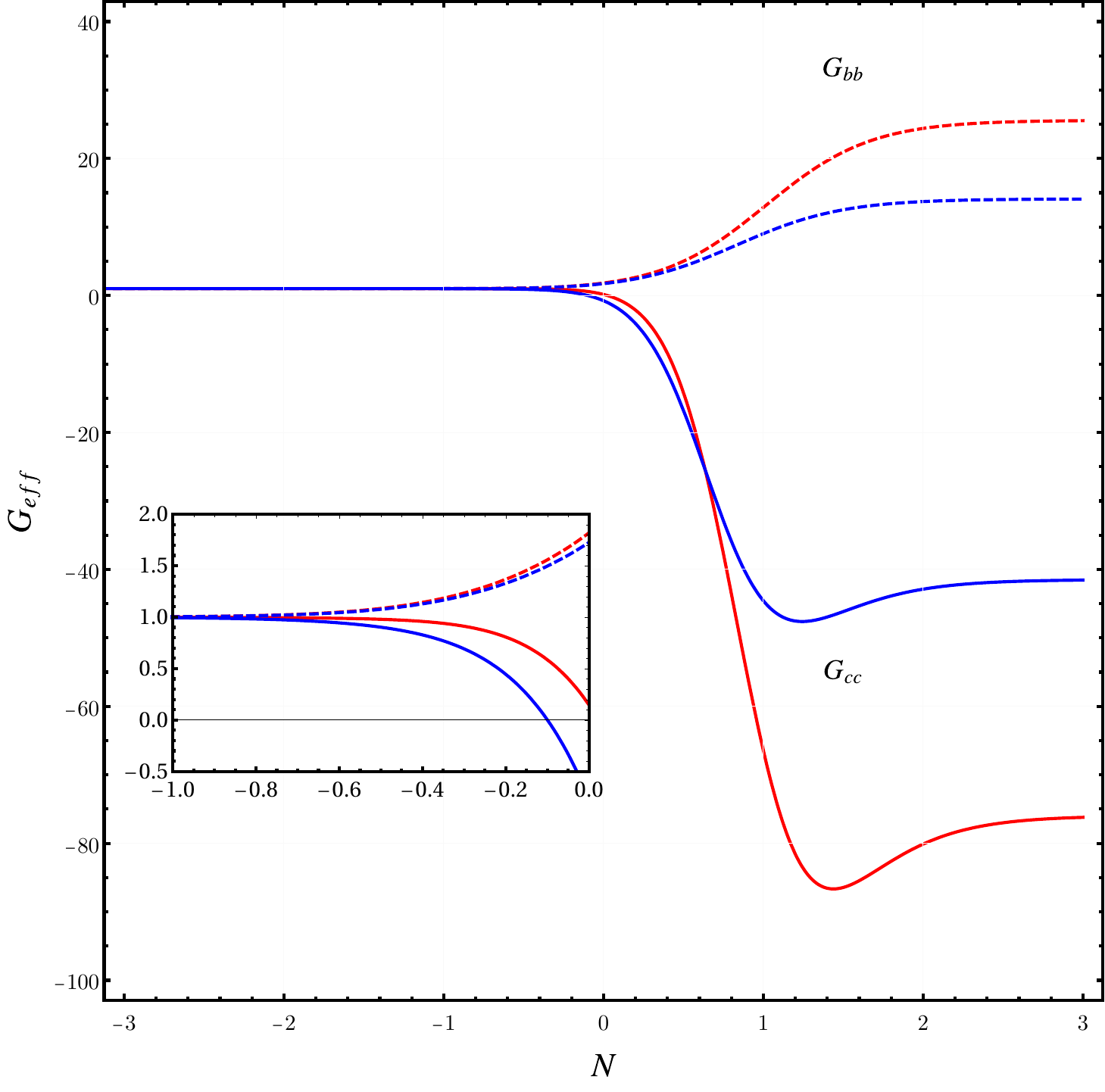}
	\caption{Left: Evolution of the EoS of DE vs $N$ (right panel).  Right:The effective gravitational constants $G_{cc}$ (solid lines), and $G_{bb}$ (dashed lines). In both panels we use the best fit parameters for ICG (in blue) and TICG (in red) for the combination BAO/CMB+H+RSD+SN.			 
	}
	\label{geff}
\end{figure*}

%%%%%%%%%%%%%%%%%%%%%%%%%%%%%%%%%%%%%%%%%%%%%%%%%%%%%
In the left panel of Fig. \ref{geff}, we plot the EoS of DE, $w_x$, for the combination of the full datasets, using the best fit values of the initial conditions $r_{i1}$ and $r_{i2}$. As illustrated in this figure, the DE EoS in TICG always remains in the phantom regime. In ICG, it undergoes a rapid transition around the epoch of matter-radiation equality, which occurs at $z_{eq}\simeq3380$ using the best fit parameters. It then stays constant at the onset of the time of recombination before undergoing a second rapid transition around $z_t \sim 1$. This evolution in the phantom regime  have strong impact on  $H_0$ and $f\sigma_8$, as has been confirmed above. This scenario, first discovered in the Covariant Galileon field \cite{DeFelice:2010pv}, is also consistent with the transitional dark energy (TDE) scenario \cite{Zhou:2021xov,Keeley:2019esp}.
In the right panel, the evolution of the effective gravitational constants $G_{cc}$ and $G_{bb}$ with redshift is plotted using the best fit parameters. We observe that for both the ICG and TICG models, the effective gravitational coupling $G_{cc}$ during the DE-dominated era is smaller than unity starting approximately from the transitional redshift at $z_t$, and even becomes negative near the present time. This latter behavior may lead to slower structural growth in TICG; however, it is significantly more pronounced in ICG, resulting in markedly reduced growth rates. In contrast, the baryon-baryon effective coupling $G_{bb}$ follows the opposite trend. 
%%%%%%%%%%%%%%%%%%%%%%%%%%%%%%%%%%%%%%%%%%%%%%%%%%%%%%%%%
%%%%%%%%%%%%%%%%%%%%%%%%%%%%%%%%%%%%%%%%%%%%%%%%%%%%%%%%%

\begin{figure}[H]
		\begin{center}
			\includegraphics[width=8cm,height=6cm]{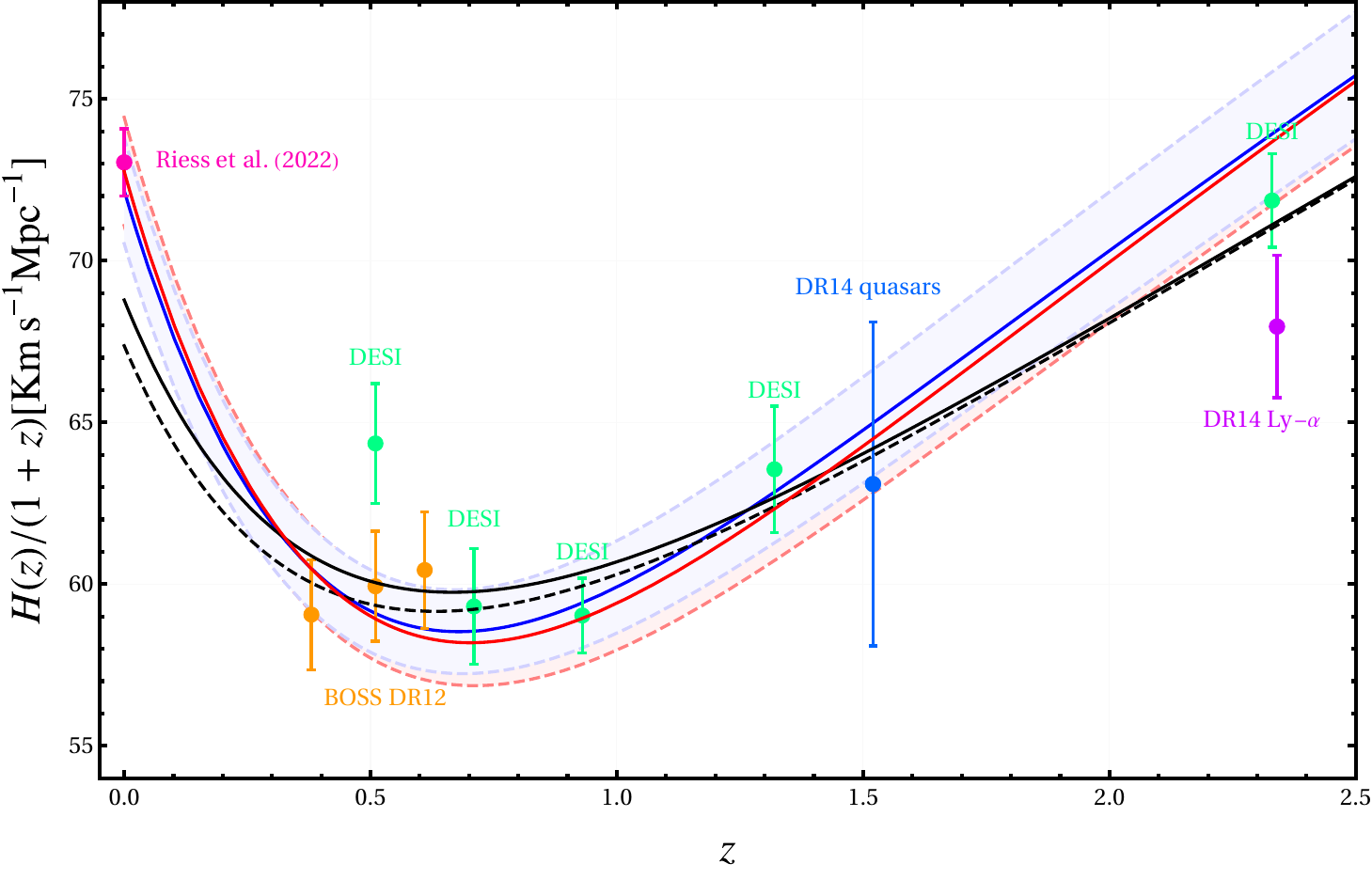}
			\caption{Redshift evolution of $H(z)/(1 + z)$ for the full combination of datasets. Data points of local measurements by Riess et al.[33] (for $H_0$), BOSS DR12 \cite{p2}, BOSS DR14 quasars \cite{2018MNRAS.477.1639Z}, BOSS DR14 Ly-$\alpha$
\cite{2019A&A629A85D,2019A&A629A86B}, and DESI 2024 \cite{desicollaboration2024desi} are shown. The Blue, red,  black and black dashed curves are for the ICG, TICG, the best fit $\Lambda\text{CDM}$ and  $\Lambda\text{CDM}$ with Planck 18,  respectively.}
			\label{fh}
		\end{center}
	\end{figure}

In Figure \ref{fh}, we present the theoretical predictions for  $H(z)/(1+z)$ derived from the combination of the full datasets. We showcase the best-fit projections for ICG, TICG, and $\Lambda$CDM models, alongside $\Lambda$CDM with Planck 18 data, juxtaposed with the DESI Collaboration measurements \cite{desicollaboration2024desi}, BOSS DR12 \cite{p2}, BOSS DR14 quasars \cite{2018MNRAS.477.1639Z}, and BOSS DR14 Lyman-$\alpha$ forest		 \cite{2019A&A629A85D,2019A&A629A86B}  data points. Notably, the ICG and TICG models demonstrate a better fit to $H(z)$ for the data   reported by DESI Collaboration, compared to $\Lambda$CDM.  The anomaly in the data point at $z=0.51$ has recently been addressed \cite{colgáin2024does}, while waiting for the future data that will be provided by the DESI Collaboration. For the remaining non-DESI data, we observe the same mediocre fit for all models. Moreover, ICG and TICG
predict a higher present day expansion rate $H_0$, evident in  the disparity between the red and blue curves compared to the black curves at $z=0$. The consistent behaviors illustrated in  Figure \ref{fh} during later times for the ICG and TICG models, beginning from the DM-DE equality, are attributed to the shared de Sitter attractor present in both models at those stages.  We also observe the crossing of the ICG and TICG curves through the $\Lambda$CDM curve at the redshift of DM-DE equality $z\sim 0.3$, and $z_t \sim 1$, where the rapid transition of the DE EoS occurs.
Let us now use an other diagnostic to asses the quality of our results regarding the latest BAO release provided by the DESI collaboration \cite{desicollaboration2024desi}. This diagnostic used the quantity defined as \cite{OColgain2022}
\begin{equation}
A(z)=\frac{D_M (z)}{z D_H (z)},
\end{equation}
where
\begin{equation}
D_M (z)=\frac{c}{H_0}\int_{0}^{z}\frac{dz'}{E(z')},\,\,D_H (z)=\frac{c}{E(z)},
\end{equation}
are the transverse and line-of-sight comoving
distances, respectively,  and $E(z)$ is the normalized Hubble parameter $H(z)=H_0 E(z)$. The notable feature of $A(z)$ lies in its independence from the sound horizon at the baryon drag epoch $r_d$. This property facilitates the computation of $A(z)$ if $E(z)$ is known.
%%%%%%%%%%%%%%%%%%%%%%%%%%%%%%%%%%%%%%%%%%%%%%%%%%%%%%%%%%ùù
\begin{figure}[t]
\centering
\includegraphics[width=8cm,height=6cm]{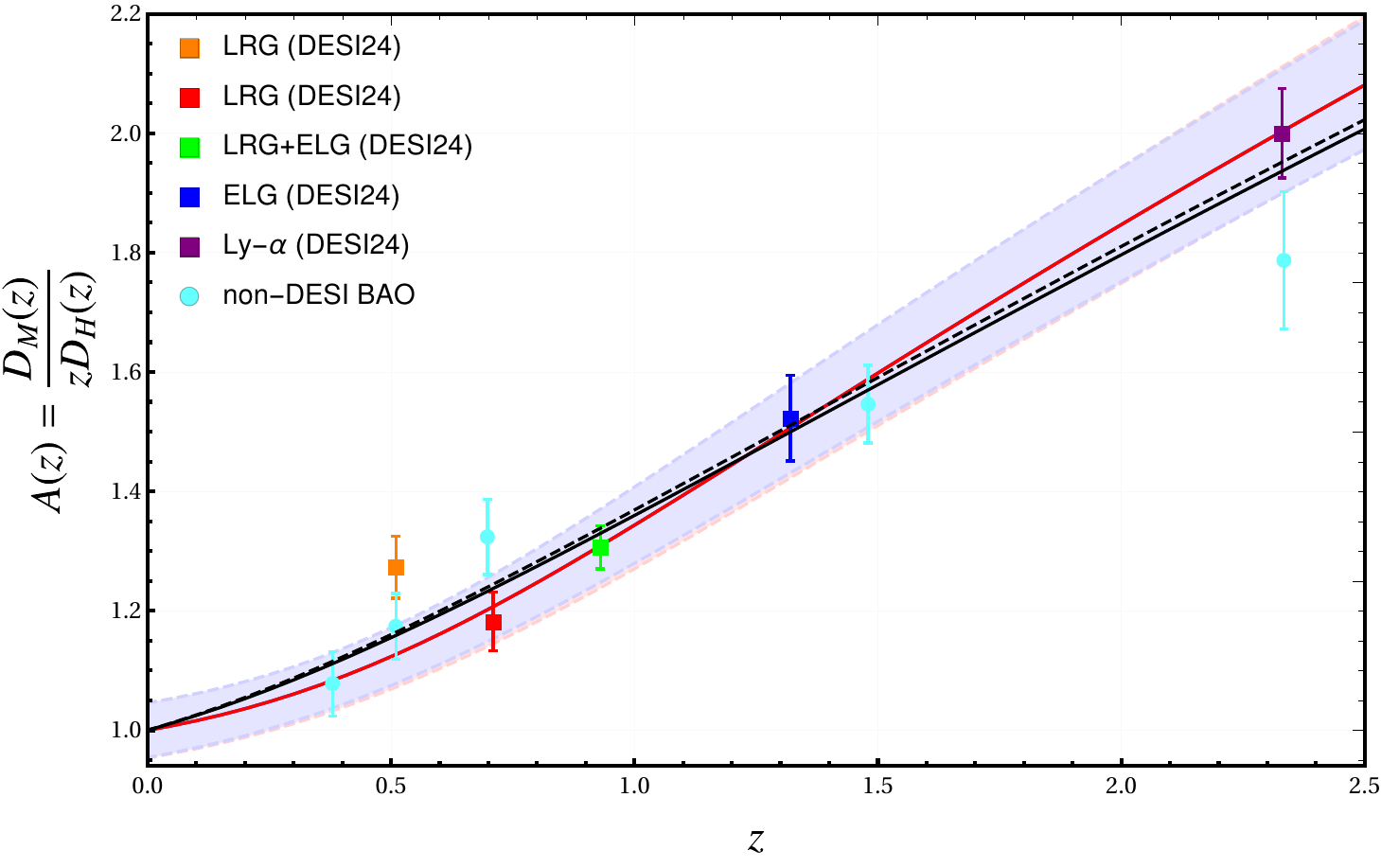}
\caption{
In Figure \ref{Proof-DESI}, we present $A(z)$ versus $z$ for the full dataset. The ICG and TICG curves (in blue and red, respectively) are indistinguishable, while the $\Lambda$CDM model  is shown with the best fit results in a solid black curve and with the Planck 18 data in a dashed black curve. Data points and their $1\sigma$ errors are listed in Tables 1 and 4 in \cite{dinda2024new}}
\label{Proof-DESI}. 
\end{figure}
%%%%%%%%%%%%%%%%%%%%%%%%%%%%%%%%%%%%%%%%%%%%%%%%%%%%%%%%%%%	
In Figure \ref{Proof-DESI}, we clearly observe that the $A(z)$ curves for the ICG and TICG models are indistinguishable and provide an excellent fit to the DESI BAO data \cite{desicollaboration2024desi}, with the notable exception of the anomalous data point at $z=0.51$. Compared to the flat $\Lambda$CDM model, we observe significant evidence for dynamical phantom regime. An other consequence of the particular behavior of the EoS of DE is reflected in the crossing of the $\Lambda$CDM curve at redshift $z\sim z_t$. The observed degeneracy between ICG and TICG concerning the $A(z)$-test suggests that this evaluation alone might not suffice for effectively discriminating between different models.
 
%%%%%%%%%%%%%%%%%%%%%%%%%%%%%%%%%%%%%%%%%%%%%%%%%%%%%%%%%%%
	\begin{figure}[H]
		\begin{center}
			\includegraphics[width=8cm,height=6cm]{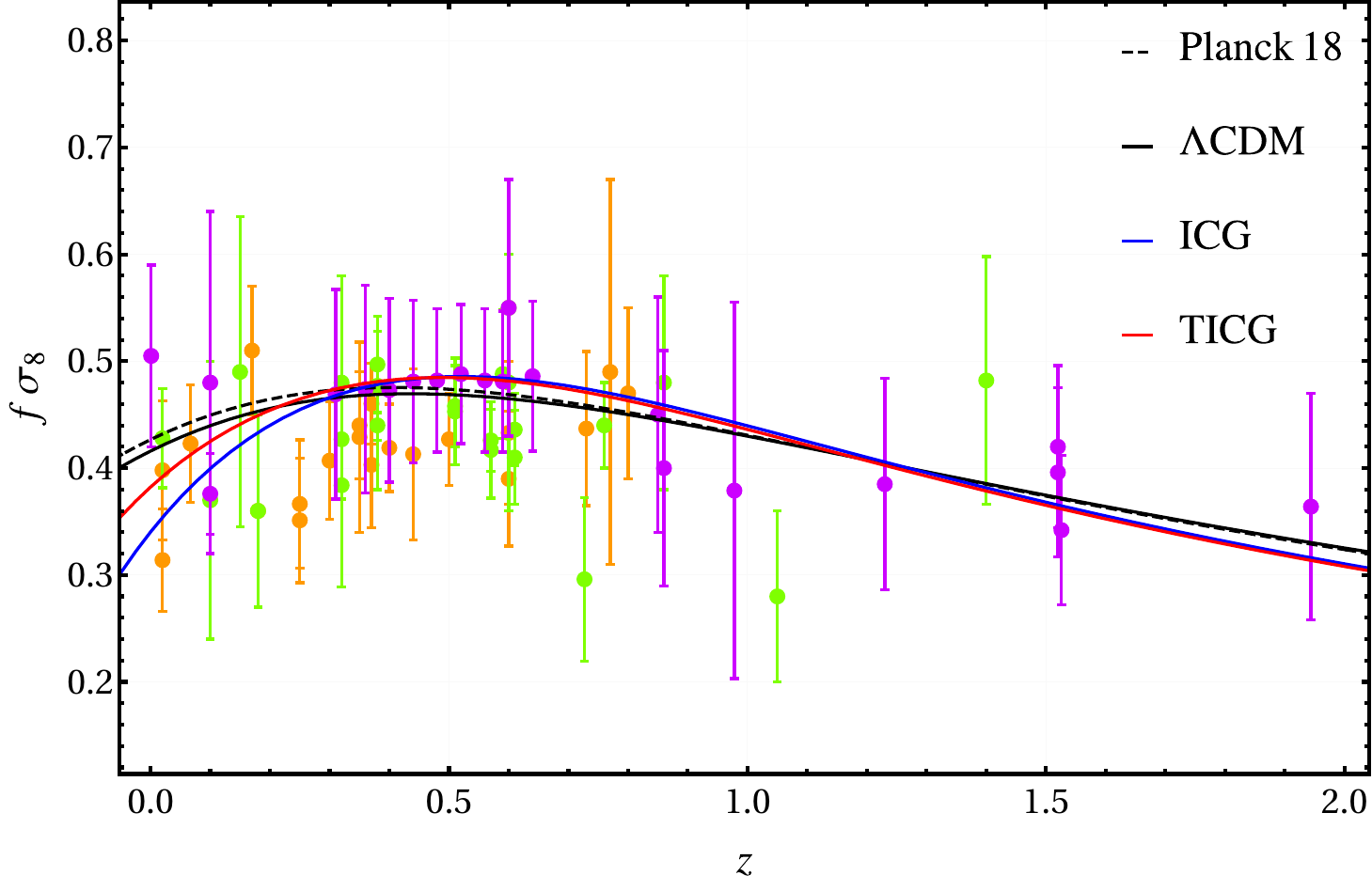}
			\caption{\label{contrast}Redshift evolution of  $f\sigma_8(z) $ for the full combination of datasets. The observational data are for the 21 latest (Violet), earliest  (Orange) and the remaining  (Green) data points taken from the Table in Appendix \ref{appendixA}. The black and the black dashed curves are the best fit $\Lambda\text{CDM}$  and  $\Lambda\text{CDM}$ with Planck 18, respectively.}
			\label{fsigma8}
		\end{center}
	\end{figure}
	
In Figure \ref{fsigma8}, we plot the best-fit behavior of $f\sigma_8$ for the ICG and TICG models, comparing them to the best-fit Planck 18 $\Lambda$CDM model. The ICG and TICG models are indistinguishable before the DM-DE equality epoch a $z \sim 0.3$, and exhibit less growth than in the $\Lambda$CDM model and Planck data in the present epoch, starting at the DM-DE  equality redshift. This behavior can be attributed to the weaker effective gravitational constant, essentially due to $G_{cc}$ for the ICG model.
An interesting pattern emerges from Figures \ref{fh} and \ref{fsigma8}. Specifically, we observe two transition redshifts at $z\sim 0.3$ and $z \sim z_t$, where the Hubble parameter and the rate of structure growth in the ICG and TICG models deviate from those in the $\Lambda$CDM model. This behavior is related to the particular evolution of the EoS of DE and the effective gravitational constants, particularly $G_{cc}$ as illustrated in figure \ref{geff}. Notably, for $ 0.3 \lesssim z \lesssim z_t $, the growth rate of structures in ICG and TICG closely matches the latest RSD measurements (depicted in violet) with remarkable precision, similar to how $H(z)$ and $A(z)$ correspond closely to measurements obtained by the DESI Collaboration \cite{desicollaboration2024desi}. We note that, even though we used only the recently published RSD data in our study, our results remain consistent with the full set of data presented in the Table in appendix \ref{appendixA}. The decline in the growth rate of structures reported by ICG is notably sharper compared to both TICG and CDM,  essentially  attributed to the negative trend of $G_{cc}$ beyond $z > 0.3$. Meanwhile, the rate of structure growth in TICG decreases moderately in contrast to $\Lambda$CDM. This latter observation elucidates why the tension on $S_8$ between Planck and LSS observations is entirely alleviated within the TICG model.

\section{Conclusions\label{sec:Conclusion}}
	
	In this study,  we explored the richness of  interacting dark energy and dark matter  cosmology in the framework of the cubic covariant Galileon model, and its efficacy in addressing the enduring and notably significant $H_0$ and $S_8$ tensions. Our investigation focused on an interaction term proportional to the Hubble parameter and Galileon dark energy, expressed as $Q=\alpha H\rho_{x}$. By initially considering the existence of a de Sitter (dS) cosmological era, we were able to express the free parameters in the Galileon Lagrangian in terms of the coupling constant, thus reducing the dimensionality of the parameter space. Utilizing appropriate dimensionless variables and transforming the field equations into an autonomous system of first-order differential equations, we conducted a comprehensive dynamical system analysis. This analysis revealed the typical physical phase space, encompassing various cosmological epochs such as radiation domination, matter domination, and dark energy domination, ultimately leading to a de Sitter expansion in the future.

The identification of a stable attractor in the dS era enabled the construction of an approximated tracking solution, closely mimicking the exact solution, particularly in the recent past before reaching the dS era. Notably, along the tracker solution, the dark energy equation of state exhibited distinctive behaviors corresponding to different cosmological epochs, showing particularly a transitional behavior as the one observed in transitional dark energy models (TDE), and first observed in the interacting covariant Galileon field.

Transitioning to the analysis of parameter constraints for the exact solution (ICG), the approximate tracker solution (TICG), and the $\Lambda$CDM model, our results provided valuable insights. Our examination of the BAO/CMB+H+RSD and BAO/CMB+H+RSD+SN combinations revealed a preference for a phantom regime in both solutions. We found that the current cosmological tensions on $H_0$ and $S_8$ are alleviated at the $68\%$ confidence level within the TICG solution. This alleviation is attributed to the observed anti-correlation between the DM-DE coupling constant $\alpha$ and the Hubble constant $H_0$, and the correlation between $\alpha$ and $S_8$. This suggests that the cosmological tensions within TICG are driven by the interaction between dark matter  and dark energy, while in ICG, the resolution of the $H_0$ tension is driven by the phantom regime.

In summary, our study advances our understanding of the complex dynamics within the interacting cubic covariant Galileon model. We emphasize the importance of the tracker solution, which emerges as a compelling avenue for addressing the $H_0$ and $S_8$ tensions. While frequentist evidence analysis favors the $\Lambda$CDM model over the ICG and TICG solutions, our findings underscore the potential of this model to provide valuable insights into some of the foremost challenges in modern cosmology.

		\acknowledgments
		{KN was financially supported by the Algerian Ministry of Higher Education and Scientific Research (MESRS).}

\bibliographystyle{JHEP} 
\bibliography{refs}
	
\appendix
\section{RSD and OHD data tables }\label{appendixA}

	%\begin{minipage}[t]{50mm}%
		\begin{table}
         \centering
		\setlength\extrarowheight{-3pt}
		\scalebox{0.7}{
		\begin{tabular}{ccccccccc}
		
		\hline
			{\scriptsize{}Index} & {\scriptsize{}Data set} & {\scriptsize{}$z$ } & {\scriptsize{}$\mathit{f\sigma_8}(z)$} & {\scriptsize{}References}&{\scriptsize{}Index} & {\scriptsize{}$z$} & {\scriptsize{}$H$} & {\scriptsize{}References}\tabularnewline
			\hline
			{\scriptsize{}1} & {\scriptsize{}SDSS-LRG} & {\scriptsize{}0.35} & {\scriptsize{}0.440 \textpm{} 0.050} & {\scriptsize{}\cite{p75}}&{\scriptsize{}1} & {\scriptsize{}0.0708} & {\scriptsize{}$69.0\pm19.68$} & {\scriptsize{}\cite{h69}}
			\tabularnewline
			{\scriptsize{}2} & {\scriptsize{}VVDS} & {\scriptsize{}0.77} & {\scriptsize{}0.490 \textpm{} 0.18} & {\scriptsize{}\cite{p75}}&{\scriptsize{}2} & {\scriptsize{}0.09} & {\scriptsize{}$69.0\pm12.0$} & {\scriptsize{}\cite{Jimenez_2002}}
			\tabularnewline
			{\scriptsize{}3} & {\scriptsize{}2dFGRS} & {\scriptsize{}0.17} & {\scriptsize{}0.510 \textpm{} 0.060} & {\scriptsize{}\cite{p75}}&{\scriptsize{}3}& {\scriptsize{}0.12} & {\scriptsize{}$68.6\pm26.2$} & {\scriptsize{}\cite{h69}}
			\tabularnewline
			{\scriptsize{}4} & {\scriptsize{}2MRS} & {\scriptsize{}0.02} & {\scriptsize{}0.314 \textpm{} 0.048 } & {\scriptsize{}\cite{p77,p78}}&{\scriptsize{}4}& {\scriptsize{}0.17} & {\scriptsize{}$83.0\pm8.0$} & {\scriptsize{}\cite{h70}}
			\tabularnewline
			{\scriptsize{}5} & {\scriptsize{}SnIa+IRAS} & {\scriptsize{}0.02} & {\scriptsize{}0.398 \textpm{} 0.065 } & {\scriptsize{}\cite{p78,p79}}&{\scriptsize{}5}& {\scriptsize{}0.179} & {\scriptsize{}$75.0\pm4.0$} & {\scriptsize{}\cite{h71}}
			\tabularnewline
			{\scriptsize{}6} & {\scriptsize{}SDSS-LRG-200} & {\scriptsize{}0.25} & {\scriptsize{}0.3512 \textpm{} 0.0583} & {\scriptsize{}\cite{p80}}&{\scriptsize{}6}& {\scriptsize{}0.199} & {\scriptsize{}$75.0\pm5.0$} & {\scriptsize{}\cite{h71}}
			\tabularnewline
			{\scriptsize{}7} & {\scriptsize{}SDSS-LRG-200} & {\scriptsize{}0.37} & {\scriptsize{}0.4602 \textpm{} 0.0378} & {\scriptsize{}\cite{p80}}&{\scriptsize{}7}& {\scriptsize{}0.20} & {\scriptsize{}$72.9\pm29.6$} & {\scriptsize{}\cite{h69}}
			\tabularnewline
			{\scriptsize{}8} & {\scriptsize{}SDSS-LRG-60 } & {\scriptsize{}0.25} & {\scriptsize{}0.3665 \textpm{} 0.0601} & {\scriptsize{}\cite{p80}}&{\scriptsize{}8}& {\scriptsize{}0.27} & {\scriptsize{}$77.0\pm14.0$} & {\scriptsize{}\cite{h70}}
			\tabularnewline
			{\scriptsize{}9} & {\scriptsize{}SDSS-LRG-60 } & {\scriptsize{}0.37} & {\scriptsize{}0.4031 \textpm{} 0.0586} & {\scriptsize{}\cite{p80}}&{\scriptsize{}9}& {\scriptsize{}0.28} & {\scriptsize{}$88.8\pm36.6$} & {\scriptsize{}\cite{h69}}
			\tabularnewline
			{\scriptsize{}10} & {{\scriptsize{}WiggleZ}} & {\scriptsize{}0.44} & {\scriptsize{}0.413 \textpm{} 0.080} & {\scriptsize{}\cite{p46}}&{\scriptsize{}10}& {\scriptsize{}0.35} & {\scriptsize{}$82.0\pm4.85$} & {\scriptsize{}\cite{h72}}
			\tabularnewline
			{\scriptsize{}11} & {\scriptsize{}WiggleZ} & {\scriptsize{}0.60} & {\scriptsize{}0.390 \textpm{} 0.063} & {\scriptsize{}\cite{p46}}&{\scriptsize{}11}& {\scriptsize{}0.352} & {\scriptsize{}$83.0\pm14.0$} & {\scriptsize{}\cite{h73}}
			\tabularnewline
			{\scriptsize{}12} & {\scriptsize{}WiggleZ} & {\scriptsize{}0.73} & {\scriptsize{}0.437 \textpm{} 0.072} & {\scriptsize{}\cite{p46}}&{\scriptsize{}12}& {\scriptsize{}0.3802} & {\scriptsize{}$83.0\pm13.5$} & {\scriptsize{}\cite{h73}}
			\tabularnewline
			{\scriptsize{}13} & {\scriptsize{}6dFGS} & {\scriptsize{}0.067} & {\scriptsize{}0.423 \textpm{} 0.055} & {\scriptsize{}\cite{p81}}&{\scriptsize{}13}& {\scriptsize{}0.4} & {\scriptsize{}$95.0\pm17.0$} & {\scriptsize{}\cite{h70}}
			\tabularnewline
			{\scriptsize{}14} & {\scriptsize{}SDSS-BOSS} & {\scriptsize{}0.30} & {\scriptsize{}0.407 \textpm{} 0.055} & {\scriptsize{}\cite{p82}}&{\scriptsize{}14}& {\scriptsize{}0.4004} & {\scriptsize{}$77.0\pm10.2$} & {\scriptsize{}\cite{h73}}
			\tabularnewline
			{\scriptsize{}15} & {\scriptsize{}SDSS-BOSS} & {\scriptsize{}0.40} & {\scriptsize{}0.419 \textpm{} 0.041} & {\scriptsize{}\cite{p82}}&{\scriptsize{}15}& {\scriptsize{}0.4247} & {\scriptsize{}$87.1\pm11.2$} & {\scriptsize{}\cite{h73}}
			\tabularnewline
			{\scriptsize{}16} & {\scriptsize{}SDSS-BOSS} & {\scriptsize{}0.50} & {\scriptsize{}0.427 \textpm{} 0.043} & {\scriptsize{}\cite{p82}}&{\scriptsize{}16}& {\scriptsize{}0.4497} & {\scriptsize{}$92.8\pm12.9$} & {\scriptsize{}\cite{h73}}
			\tabularnewline
			{\scriptsize{}17} & {\scriptsize{}SDSS-BOSS} & {\scriptsize{}0.60} & {\scriptsize{}0.433 \textpm{} 0.067} & {\scriptsize{}\cite{p82}}&{\scriptsize{}17}& {\scriptsize{}0.4783} & {\scriptsize{}$80.9\pm9.0$} & {\scriptsize{}\cite{h73}}
			\tabularnewline
			{\scriptsize{}18} & {\scriptsize{}Vipers} & {\scriptsize{}0.80} & {\scriptsize{}0.470 \textpm{} 0.080} & {\scriptsize{}\cite{p83}}&{\scriptsize{}18}& {\scriptsize{}0.48} & {\scriptsize{}$97.0\pm62.0$} & {\scriptsize{}\cite{h74}}
			\tabularnewline
			{\scriptsize{}19} & {\scriptsize{}SDSS-DR7-LRG} & {\scriptsize{}0.35} & {\scriptsize{}0.429 \textpm{} 0.089} & {\scriptsize{}\cite{p84}}&{\scriptsize{}19}& {\scriptsize{}0.593} & {\scriptsize{}$104.0\pm13.0$} & {\scriptsize{}\cite{h71}}
			\tabularnewline
			{\scriptsize{}20} & {\scriptsize{}GAMA} & {\scriptsize{}0.18} & {\scriptsize{}0.360 \textpm{} 0.090} & {\scriptsize{}\cite{p86}}&{\scriptsize{}20}& {\scriptsize{}0.68} & {\scriptsize{}$92.0\pm8.0$} & {\scriptsize{}\cite{h71}}
			\tabularnewline
			{\scriptsize{}21} & {\scriptsize{}GAMA} & {\scriptsize{}0.38} & {\scriptsize{}0.440 \textpm{} 0.060} & {\scriptsize{}\cite{p86}}&{\scriptsize{}21}& {\scriptsize{}0.781} & {\scriptsize{}$105.0\pm12.0$} & {\scriptsize{}\cite{h71}}
			\tabularnewline
			{\scriptsize{}22} & {\scriptsize{}BOSS-LOWZ} & {\scriptsize{}0.32} & {\scriptsize{}0.384 \textpm{} 0.095} & {\scriptsize{}\cite{p87}}&{\scriptsize{}22}& {\scriptsize{}0.875} & {\scriptsize{}$125.0\pm17.0$} & {\scriptsize{}\cite{h71}}
			\tabularnewline
			{\scriptsize{}23} & {\scriptsize{}SDSS DR10 and DR11 } & {\scriptsize{}0.32} & {\scriptsize{}0.48 \textpm{} 0.10} & {\scriptsize{}\cite{p87}}&{\scriptsize{}23}& {\scriptsize{}0.88} & {\scriptsize{}$90.0\pm40.0$} & {\scriptsize{}\cite{h74}}
			\tabularnewline
			{\scriptsize{}24} & {\scriptsize{}SDSS DR10 and DR11 } & {\scriptsize{}0.57} & {\scriptsize{}0.417 \textpm{} 0.045} & {\scriptsize{}\cite{p87}}&{\scriptsize{}24}& {\scriptsize{}0.9} & {\scriptsize{}$117.0\pm23.0$} & {\scriptsize{}\cite{h70}}
			\tabularnewline
			{\scriptsize{}25} & {\scriptsize{}SDSS-MGS} & {\scriptsize{}0.15} & {\scriptsize{}0.490 \textpm{} 0.145} & {\scriptsize{}\cite{p89}}&{\scriptsize{}25}& {\scriptsize{}1.037} & {\scriptsize{}$154.0\pm12.0$} & {\scriptsize{}\cite{h71}}
			\tabularnewline
			{\scriptsize{}26} & {\scriptsize{}SDSS-veloc} & {\scriptsize{}0.10} & {\scriptsize{}~0.370 \textpm{} 0.130~} & {\scriptsize{}\cite{p90}}&{\scriptsize{}26}& {\scriptsize{}1.3} & {\scriptsize{}$168.0\pm17.0$} & {\scriptsize{}\cite{h70}}
			\tabularnewline
			{\scriptsize{}27} & {\scriptsize{}FastSound} & {\scriptsize{}1.40} & {\scriptsize{}0.482 \textpm{} 0.116} & {\scriptsize{}\cite{p92}}&{\scriptsize{}27}& {\scriptsize{}1.363} & {\scriptsize{}$160.0\pm33.6$} & {\scriptsize{}\cite{h75}}
			\tabularnewline
			{\scriptsize{}28} & {\scriptsize{}SDSS-CMASS} & {\scriptsize{}0.59} & {\scriptsize{}0.488 \textpm{} 0.060} & {\scriptsize{}\cite{p94}}&{\scriptsize{}28}& {\scriptsize{}1.43} & {\scriptsize{}$177.0\pm18.0$} & {\scriptsize{}\cite{h70}}
			\tabularnewline
			{\scriptsize{}29} & {\scriptsize{}BOSS DR12} & {\scriptsize{}0.38} & {\scriptsize{}0.497 \textpm{} 0.045} & {\scriptsize{}\cite{p2}} &{\scriptsize{}29}& {\scriptsize{}1.53} & {\scriptsize{}$140.0\pm14.0$} & {\scriptsize{}\cite{h70}}
			\tabularnewline
			{\scriptsize{}30} & {\scriptsize{}BOSS DR12} & {\scriptsize{}0.51} & {\scriptsize{}0.458 \textpm{} 0.038 } & {\scriptsize{}\cite{p2}}&{\scriptsize{}30}& {\scriptsize{}1.75} & {\scriptsize{}$202.0\pm40.0$} & {\scriptsize{}\cite{h70}}
			\tabularnewline
			{\scriptsize{}31} & {\scriptsize{}BOSS DR12} & {\scriptsize{}0.61} & {\scriptsize{}0.436 \textpm{} 0.034} & {\scriptsize{}\cite{p2}}&{\scriptsize{}31}& {\scriptsize{}1.965} & {\scriptsize{}$186.5\pm50.4$} & {\scriptsize{}\cite{h75}}
			\tabularnewline
			{\scriptsize{}32} & {\scriptsize{}BOSS DR12} & {\scriptsize{}0.38} & {\scriptsize{}0.477 \textpm{} 0.051} & {\scriptsize{}\cite{p95}}\tabularnewline
			{\scriptsize{}33} & {\scriptsize{}BOSS DR12} & {\scriptsize{}0.51} & {\scriptsize{}0.453 \textpm{} 0.050} & {\scriptsize{}\cite{p95}}\tabularnewline
			{\scriptsize{}34} & {\scriptsize{}BOSS DR12} & {\scriptsize{}0.61} & {\scriptsize{}0.410 \textpm{} 0.044} & {\scriptsize{}\cite{p95}}\tabularnewline
			{\scriptsize{}35} & {\scriptsize{}Vipers v7} & {\scriptsize{}0.76} & {\scriptsize{}0.440 \textpm{} 0.040} & {\scriptsize{}\cite{p55}}\tabularnewline
			{\scriptsize{}36} & {\scriptsize{}Vipers v7} & {\scriptsize{}1.05} & {\scriptsize{}0.280 \textpm{} 0.080} & {\scriptsize{}\cite{p55}}\tabularnewline
			{\scriptsize{}37} & {\scriptsize{}BOSS LOWZ } & {\scriptsize{}0.32} & {\scriptsize{}0.427 \textpm{} 0.056} & {\scriptsize{}\cite{p96}}\tabularnewline
			{\scriptsize{}38} & {\scriptsize{}BOSS CMASS} & {\scriptsize{}0.57} & {\scriptsize{}0.426 \textpm{} 0.029} & {\scriptsize{}\cite{p96}}\tabularnewline
			{\scriptsize{}39} & {\scriptsize{}Vipers} & {\scriptsize{}0.727} & {\scriptsize{}0.296 \textpm{} 0.0765} & {\scriptsize{}\cite{p97}}\tabularnewline
			{\scriptsize{}40} & {\scriptsize{}6dFGS+SnIa} & {\scriptsize{}0.02} & {\scriptsize{}0.428 \textpm{} 0.0465} & {\scriptsize{}\cite{p98}}\tabularnewline
			{\scriptsize{}41} & {\scriptsize{}Vipers} & {\scriptsize{}0.6} & {\scriptsize{}0.48 \textpm{} 0.12} & {\scriptsize{}\cite{p99}}\tabularnewline
			{\scriptsize{}42} & {\scriptsize{}Vipers} & {\scriptsize{}0.86} & {\scriptsize{}0.48 \textpm{} 0.10 } & {\scriptsize{}\cite{p99}}\tabularnewline
			{\scriptsize{}43} & {\scriptsize{}Vipers PDR-2} & {\scriptsize{}0.60} & {\scriptsize{}0.550 \textpm{} 0.120} & {\scriptsize{}\cite{p100}}\tabularnewline
			{\scriptsize{}44} & {\scriptsize{}Vipers PDR-2} & {\scriptsize{}0.86} & {\scriptsize{}0.400 \textpm{} 0.110} & {\scriptsize{}\cite{p100}}\tabularnewline
			{\scriptsize{}45} & {\scriptsize{}SDSS DR13} & {\scriptsize{}0.1} & {\scriptsize{}0.48 \textpm{} 0.16} & {\scriptsize{}\cite{p101}}\tabularnewline
			{\scriptsize{}46} & {\scriptsize{}2MTF} & {\scriptsize{}0.001} & {\scriptsize{}0.505 \textpm{} 0.085} & {\scriptsize{}\cite{p102}}\tabularnewline
			{\scriptsize{}47} & {\scriptsize{}Vipers PDR-2} & {\scriptsize{}0.85} & {\scriptsize{}0.45 \textpm{} 0.11} & {\scriptsize{}\cite{p103}}\tabularnewline
			{\scriptsize{}48} & {\scriptsize{}BOSS DR12} & {\scriptsize{}0.31} & {\scriptsize{}0.469 \textpm{} 0.098} & {\scriptsize{}\cite{p49}}\tabularnewline
			{\scriptsize{}49} & {\scriptsize{}BOSS DR12} & {\scriptsize{}0.36} & {\scriptsize{}0.474 \textpm{} 0.097} & {\scriptsize{}\cite{p49}}\tabularnewline
			{\scriptsize{}50} & {\scriptsize{}BOSS DR12} & {\scriptsize{}0.40} & {\scriptsize{}0.473 \textpm{} 0.086} & {\scriptsize{}\cite{p49}}\tabularnewline
			{\scriptsize{}51} & {\scriptsize{}BOSS DR12} & {\scriptsize{}0.44} & {\scriptsize{}0.481 \textpm{} 0.076} & {\scriptsize{}\cite{p49}}\tabularnewline
			{\scriptsize{}52} & {\scriptsize{}BOSS DR12} & {\scriptsize{}0.48} & {\scriptsize{}0.482 \textpm{} 0.067} & {\scriptsize{}\cite{p49}}\tabularnewline
			{\scriptsize{}53} & {\scriptsize{}BOSS DR12} & {\scriptsize{}0.52} & {\scriptsize{}0.488 \textpm{} 0.065} & {\scriptsize{}\cite{p49}}\tabularnewline
			{\scriptsize{}54} & {\scriptsize{}BOSS DR12} & {\scriptsize{}0.56} & {\scriptsize{}0.482 \textpm{} 0.067} & {\scriptsize{}\cite{p49}}\tabularnewline
			{\scriptsize{}55} & {\scriptsize{}BOSS DR12} & {\scriptsize{}0.59} & {\scriptsize{}0.481 \textpm{} 0.066} & {\scriptsize{}\cite{p49}}\tabularnewline
			{\scriptsize{}56} & {\scriptsize{}BOSS DR12} & {\scriptsize{}0.64} & {\scriptsize{}0.486 \textpm{} 0.070} & {\scriptsize{}\cite{p49}}\tabularnewline
			{\scriptsize{}57} & {\scriptsize{}SDSS DR7} & {\scriptsize{}0.1} & {\scriptsize{}0.376 \textpm{} 0.038} & {\scriptsize{}\cite{p104}}\tabularnewline
			{\scriptsize{}58} & {\scriptsize{}SDSS-IV} & {\scriptsize{}1.52} & {\scriptsize{}0.420 \textpm{} 0.076} & {\scriptsize{}\cite{p105}}\tabularnewline
			{\scriptsize{}59} & {\scriptsize{}SDSS-IV} & {\scriptsize{}1.52} & {\scriptsize{}0.396 \textpm{} 0.079} & {\scriptsize{}\cite{p106}}\tabularnewline
			{\scriptsize{}60} & {\scriptsize{}SDSS-IV} & {\scriptsize{}0.978} & {\scriptsize{}0.379 \textpm{} 0.176} & {\scriptsize{}\cite{p107}}\tabularnewline
			{\scriptsize{}61} & {\scriptsize{}SDSS-IV} & {\scriptsize{}1.23} & {\scriptsize{}0.385 \textpm{} 0.099} & {\scriptsize{}\cite{p107}}\tabularnewline
			{\scriptsize{}62} & {\scriptsize{}SDSS-IV} & {\scriptsize{}1.526} & {\scriptsize{}0.342 \textpm{} 0.070} & {\scriptsize{}\cite{p107}}\tabularnewline
			{\scriptsize{}63} & {\scriptsize{}SDSS-IV} & {\scriptsize{}1.944} & {\scriptsize{}0.364 \textpm{} 0.106} & {\scriptsize{}\cite{p107}}\tabularnewline
			\hline
\end{tabular}
\label{RSD}}
\end{table}

\end{document}